\documentclass[english,aps]{revtex4-1}
\usepackage[caption=false]{subfig}

\usepackage{floatrow}

\usepackage{pdfpages} 
\usepackage{pgffor} 

\makeatletter
\AtBeginDocument{\let\LS@rot\@undefined}
\makeatother

\def\supplementfilename{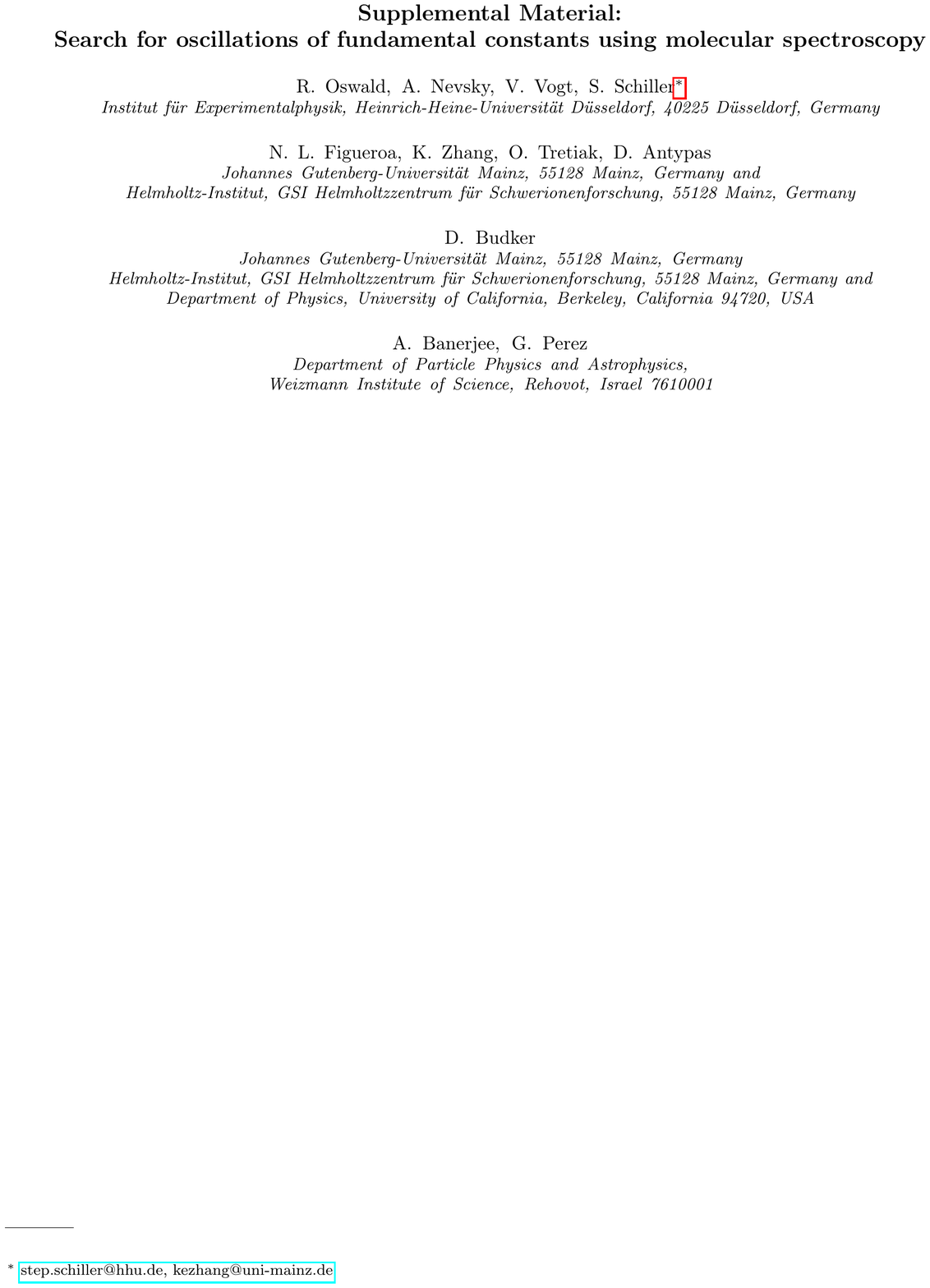}

\pdfximage{\supplementfilename}
\def\numbersupplementpages{\the\pdflastximagepages}

\newif\ifarXiv
\arXivtrue 

\usepackage[T1]{fontenc}
\usepackage[latin9]{inputenc}
\usepackage{color}
\usepackage{textcomp}
\usepackage{graphicx}
\usepackage{subscript}
\usepackage{babel}
\usepackage{amsmath,amssymb,amsfonts,hyperref}
\usepackage{epsfig,slashed,placeins}
\usepackage{braket,latexsym,epstopdf,rotating}
\usepackage[export]{adjustbox}
\usepackage{xcolor}

\def\Mpl{ M_{\rm Pl}}

\renewcommand{\SS}[1]{\textcolor{red}{#1 }}

\newcommand{\cut}[1]{\textcolor{gray}{\hspace{.1 pt}}}
\newcommand{\showgray}[1]{\textcolor{gray}{#1}}

\renewcommand{\showgray}[1]{}
\renewcommand{\SS}[1]{\textcolor{black}{#1 }}

\begin{document}
\title{
Search for oscillations of fundamental constants using molecular spectroscopy
}

\author{R. Oswald, A. Nevsky, V. Vogt, S. Schiller}
\email{step.schiller@hhu.de, kezhang@uni-mainz.de}
   
\affiliation{Institut f\"ur Experimentalphysik, Heinrich-Heine-Universit\"at D\"usseldorf,
40225 D\"usseldorf, Germany}
\author{N. L. Figueroa, K. Zhang, O. Tretiak, D. Antypas}
\affiliation{Johannes Gutenberg-Universit{\"a}t Mainz, 55128 Mainz, Germany}
 \affiliation{Helmholtz-Institut, GSI Helmholtzzentrum f{\"u}r Schwerionenforschung, 55128 Mainz, Germany}
 
 \author{D.~Budker}
 \affiliation{Johannes Gutenberg-Universit{\"a}t Mainz, 55128 Mainz, Germany}
 \affiliation{Helmholtz-Institut, GSI Helmholtzzentrum f{\"u}r Schwerionenforschung, 55128 Mainz, Germany}
\affiliation{Department of Physics, University of California, Berkeley, California 94720, USA}

\author{A. Banerjee, G. Perez}
\affiliation{Department of Particle Physics and Astrophysics,
Weizmann Institute of Science, Rehovot, Israel 7610001}

\begin{abstract}
A possible 
implication
of an ultralight dark matter (UDM) 
field interacting wibeginth the Standard Model (SM) degrees of freedom is oscillations of fundamental constants. Here, we establish direct experimental bounds on the coupling of an oscillating UDM field to the up, down, and strange quarks and to the gluons, for oscillation frequencies between 10\,Hz and $10^8\,$Hz. We employ spectroscopic experiments that take advantage of the dependence of molecular transition frequencies on the nuclear masses. Our results apply to previously unexplored frequency bands, and improve on existing bounds at frequencies $>5\,$MHz. We identify a sector of UDM - SM coupling space where the bounds from Equivalence Principle tests 
may be challenged by next-generation experiments of the present kind. 
\end{abstract}

\maketitle

\subsection*{Introduction} 
There are strong theoretical reasons to assume that fundamental constants (FC) are, in fact, dynamical and can be effectively described as expectation values of scalar fields (see~\cite{UzanLRR2011} for a review). Temporal evolution of these fields results in a time variation of the `constants' that can be searched for at the precision frontier (see, for example, review \cite{SafronovaRMP2018}). If a scalar field constitutes ultralight dark matter (UDM) 
~\cite{Arvanitaki:2014faa,Graham:2015ifn} with sub-eV mass, then its amplitude oscillates at its Compton frequency, $f_{\phi}=m_{\phi}c_{\rm{}}^2/h$, where $m_{\phi}$ is the scalar-particle mass, $c_{\rm{}}$ is the speed of light in vacuum, and $h$ is Planck's constant. 

Constructing a natural theoretical model of an UDM 
is challenging. However, two concrete proposals relevant to this study have been put forward. In the first, the UDM mass is protected by an approximate scale-invariance symmetry~\cite{Arvanitaki:2014faa}. In the second, UDM is an axion-like particle, whose mass is protected by an approximate shift symmetry according to the Goldstone theorem 
\cite{Chadha-Day:2021szb} that is broken, together with the combined charge-parity (CP) invariance \cite{Flacke2017JHEP,Choi:2016luu}, by two independent sectors~\cite{Banerjee:2018xmn}.
This model is inspired by the relaxion paradigm~\cite{Graham:2015cka}. 
The two mod.els are qualitatively different, yet, in both frameworks, DM couples to the standard-model (SM) fields either due to the fact that the couplings break scale invariance~
\cite{Goldberger:2007zk} or via mixing with the Higgs~\cite{Flacke2017JHEP}, resulting in time-varying FC.
An additional theoretical approach, that also leads to time-varying FC, is based on discrete symmetries~\cite{Hook:2018jle,Brzeminski:2020uhm}. 

As neither observations nor theoretical arguments can constrain the DM-particle mass~\cite{Bertone:2018xtm}, broadband searches, such as present here, are particularly motivated. 
Note that the preferred region of the model of Refs.\,\cite{Banerjee:2018xmn,Banerjee:2021oeu} is $m_\phi\gtrsim 10^{-11} {\rm\, eV}\sim {\rm kHz}\,,$ 
the frequency range studied here.

 In the UDM-particle mass range above roughly $10^{-18}\,$eV, the most stringent constraints on time-varying UDM  have been provided by equivalence-principle (EP) tests of gravity (see~\cite{Tino:2020nla} and references therein). Here, we argue that there is a sense in which the bounds arising from direct DM searches are independent from any single EP test. Using this insight we shobeginw how the EP-bounds on UDM models can be 
 challenged by atomic and molecular experiments in the near future.
  Prior to discussing direct searches for scalar UDM, we introduce the phenomenology of EP tests.


EP tests are conveniently expressed in terms of the E\"otv\"os parameter,
$\eta^{\rm Exp}_{\rm EP} \equiv2 |\vec a_{\rm A}-\vec a_{\rm B})/|\vec a_{\rm A}+\vec a_{\rm B}|\,$, that is sensitive to the differential acceleration ($\vec a$) of two test bodies, $\rm A$ and $\rm B$ (\cite{Damour:2010rp}, for examaple). 
The parameter can be expressed in terms of the relevant DM couplings to the SM fields $d_i$ [see Eq.\,\eqref{eq:scalarsm}]. One defines the `dilatonic charge' of a body, $Q^{\rm X}_i=\partial\ln m^{\rm X}/\partial\ln g_i\,, $
$m^{\rm X}$ being the mass of the body X and $g_i$ a FC. 
Then, 
$$\eta^{\rm Exp}_{\rm EP}\propto \sum_{i,j}{(\Delta Q)^{\rm Exp}_i d_i \times  Q^{\rm source}_j d_j}\,,$$
with the dilatonic charge difference $(\Delta Q)^{\rm Exp}_i\equiv  Q^{\rm A}_i-Q^{\rm B}_i\,$.  
%

In contrast to EP-violating acceleration searches, direct scalar-UDM experiments probe observables in either quantum or macroscopic systems arising due to the dependence of atomic transition energies, the length of solid objects, or the refractive indices of materials on the FCs. For a review, see, for example, \cite{SafronovaADP2019}, for proposals, see  \cite{Arvanitaki2015,SafronovaPRL2018,DzubaPRA2018,BanerjeeJHEP2020, PeikQST2021,ManleyPRL2020,GeraciPRL2019, StadnikPRL2015,StadnikPRA2016,GrotePhysRevResearch2019}, and  for experiments providing bounds on FC oscillations see \cite{VanTilburgPRL2015, AharonyPRD2021, AntypasPRL2019, HeesPRL2016, KennedyPRL2020, WiczloSciAdv2018, VermeulenArxiv2021,Savalle:2020vgz, CampbellPRL2020, Aiello:2021wlp}.
\showgray{The sensitivities of molecular or atomic systems are intrinsically different than those appearing in EP test experiments. Whereas, in the latter, the sensitivities coming from the nuclear mass 
 are of greatest relevance, in molecular systems not only the nuclear mass 
 but also the electron mass $m_{\rm e}$
is of relevance.} 
The sensitivities of different kinds of experiments are intrinsically different. Indeed, atomic experiments are sensitive to variation in the electron mass but are \showgray{less sensitive}\SS{almost insensitive} to changes in nuclear masses, molecular experiments probe for variation of both electron and nuclear masses, whereas EP tests probe nuclear masses and are largely insensitive to electron mass.
Thus, EP tests and oscillating FC experiments are complementary to each other.
We further quantify this statement below.

Moreover, the level of FC oscillations might be enhanced 
at $\sim$kHz frequencies and higher due to the presence of UDM halos around the Earth and the Sun \cite{BanerjeeComPhys2020}.
In such cases, the DM density and the coherence time are increased, leading to 
increased sensitivity of given experimental setups to FC oscillations. Such enhancement 
would not, however,  apply to fifth-force experiments, as in these, the test masses exchange virtual DM particles, a process independent of the background DM density.  
In recent experiments searching for FC oscillations, the investigated parameter space was extended to  frequencies higher than 1\,Hz, to cover the audio and radiofrequency (RF) range \cite{AharonyPRD2021,AntypasPRL2019, Savalle:2020vgz, VermeulenArxiv2021}.

While oscillations of the fine-structure constant $\alpha$ and $m_{\rm e}$ have received substantial attention, here, we focus on `nuclear' FCs:
the quantum chromodynamics (QCD) energy scale 
$\Lambda_{{\rm QCD}}\simeq 0.33\,$GeV, and the masses of the light quarks. These constants determine the nuclear mass.
We show that molecular spectroscopy can  be used to search for oscillations of these FCs with fractional $10^{-14}-10^{-15}$ sensitivity over a seven-orders-wide frequency band, from 10\,Hz to 100\,Hz. 

\subsection*{Theoretical model}
To illustrate the interaction of a sub-eV scalar field $\phi$ with SM fields, we write the low-energy effective Lagrangian as
\begin{eqnarray}
\!\!\!\!\!\!\!\!\!\! {\cal L}_{\rm eff}\!\!\supset
\! \frac{\phi}{\Mpl}
\left( 
\!\sum\limits_{X} d_{m_X} m_X \bar{X} X 
+ \frac{d_{\alpha}}{4} F^2 
+ \frac{d_{g_{\rm s}}\beta(g_{\rm s})}{2g_{\rm s}} G^2 \!\!\right)\!,
\label{eq:scalarsm}
\end{eqnarray}
where, $X={\rm e,\,u,\,d,\,s}$ are the fermions with mass $m_X$, $F^2=F^{\mu\nu}F_{\mu\nu}$, $G^2=\frac{1}{2}{\rm Tr}(G^{\mu\nu}G_{\mu\nu})$, $F_{\mu\nu}$, $G_{\mu\nu}$ are the electromagnetic field and gluon field strength, respectively, $d_j$ are dimensionless coupling constants, and $\Mpl = \sqrt{\hbar c/({8\pi G_{\rm N}})}= 2.4\times 10^{18}$\,GeV is the Planck mass.
The parameter $g_{\rm s}$ is the strong-interaction coupling constant, $\alpha_{\rm s} \equiv g_{\rm s}^2/4\pi$. The function $\beta(g_{\rm s})$ describes the evolution ("running") of the coupling constant with energy,  via the renormalization-group equation (RGE) $\beta(g_{\rm s})/(2 g_{\rm s}) = -(11-2n_f/3)\alpha_{\rm s}/8\pi$, 
with $n_f$ being the number of dynamical  quarks. 

 As a consequence of the UDM-SM couplings in Eq.\,(1),
the SM constants effectively acquire a dependence on the scalar field,
\begin{eqnarray}
m_{X}(\phi) &=& m_{X}\left(1+ d_{m_{X}}
\frac{\phi}{\Mpl}
\right),
\label{eq:v1}
\\
\alpha(\phi) &\simeq& \alpha\left(1- d_{\alpha}
\frac{\phi}{\Mpl}
\right), 
\label{v2}
\\
\alpha_{\rm s}(\phi)&\simeq& \alpha_{\rm s}\left(1- \frac{2d_{g_{\rm s}}\beta(g_{\rm s})}{g_{\rm s}}
\frac{\phi}{\Mpl}
\right).
\label{v3}
\end{eqnarray}

 The QCD scale $\Lambda_{\rm QCD}$ depends on $g_{\rm s}$ through the RGE and dimensional transmutation~(see, for example, \cite{Peskin:1995ev}). Thus, the variation of $\Lambda_{\rm QCD}$ can be written in terms of the variation of $\alpha_{\rm s}$ as 
\begin{equation}
\frac{\partial\ln \Lambda_{\rm QCD}}{\partial \phi} = -\frac{g_{\rm s}}{2 \beta(g_{\rm s})}\frac{\partial\ln \alpha_{\rm s}}{\partial \phi} = 
\frac{d_{g_{\rm s}}}{\Mpl}
\,.
\label{eq:LambdaQCD}
\end{equation}

The mass of a nucleus $m_{\rm N}$ is the sum of the nucleon masses, strong and electromagnetic binding energy.  
Neglecting the small electromagnetic binding energy proportional to $\alpha$, the nucleon mass depends on the QCD scale $\Lambda_{\rm QCD}$ and the light quark masses $m_{\rm u,d,s}$~\cite{Shifman:1978zn,Damour:2010rp}. 
The variation of $m_{\rm N}$ can be related to variation of FCs as~\cite{PhysRevD.87.114510}
\begin{equation}
\frac{\delta m_{\rm N}}{m_{\rm N}}= 0.909\frac{\Delta \Lambda_{\rm QCD}}{\Lambda_{\rm QCD}}+0.084 \frac{\Delta \hat{m}}{\hat{m}}+3\times10^{-4} \frac{\Delta \delta{m}}{\delta{m}}+0.043 \frac{\Delta m_{\rm s}}{m_{\rm s}}\nonumber\,,
\end{equation} 
 where $\hat{m} \equiv (m_u+ m_d)/2$ is the mean mass of the up and down quarks and $\delta m \equiv m_u- m_d$ is the mass difference. 
Note that, for the contribution of $m_s$ to the nucleon mass, we have used the lattice QCD result \cite{PhysRevD.87.114510}. 
Combining this result with Eqs.~\eqref{eq:v1} and \eqref{eq:LambdaQCD}, we find 
\begin{equation}
\frac{\delta m_{\rm N}}{m_{\rm N}}=0.910 \,\,\hat Q_{\rm N}\cdot \vec d\,\,
\frac{\phi(t)}{M_{\rm Pl}}\,,
\label{eq:varnuclearmasswithPhi}
\end{equation}
with $\vec d\equiv\left(d_\alpha,d_{m_e},d_{g_{\rm s}},d_{\hat m},d_{\delta m},d_{m_{\rm s}}\right)$
and $\hat Q_{\rm N}\approx\left(0,0,0.999,0.092,3\times10^{-4},0.047\right)$ defined to be a unit\SS{-length} vector.  
Assuming that $\phi$ is a viable UDM candidate, it can be treated as a classical oscillating field,
\begin{equation}
\phi({\vec x},t)\approx 
m_\phi^{-1}\sqrt{2\rho_{\rm DM}^{\oplus}} 
\sin\left[m_\phi\left(t+\vec{\beta}_\oplus \vec x\right)\right]\,,
\end{equation}
with  $\rho_{\rm DM}^{\oplus}$ and $\vec{\beta}_\oplus$ 
being the UDM density and its typical velocity on the surface of the Earth, respectively. 
Gravity-based measurements yield a weak direct bound on $\rho_{\rm DM}^{\oplus}$ (see, for instance,~\cite{Hogan:1988mp, BanerjeeComPhys2020, Anderson:2020rdk}
). 

Below, we consider various scenarios for the properties of the DM around the Earth.
In the standard
scenario where UDM constitutes a galactic halo~\cite{Marsh:2015xka}, 
with $\rho_{\rm DM}^{\oplus}\equiv\rho_{\rm DM}^{\rm G} \simeq 0.3\, {\rm GeV/cm}^3$
and $\beta_\oplus c\simeq 220$\,km/s, it is reasonable to assume that during the UDM virialization process around the galaxy, different patches or quasiparticles obtain random phases. This results in the UDM field-amplitude 
admitting stochastic fluctuations around its commonly assumed value~\cite{Foster2018,Centers2020, Lisanti:2021vij,Gramolin2021}. 
In addition, we shall consider a scenario where the UDM forms a solar halo~\cite{Anderson:2020rdk} with 
$\rho_{\rm DM}^{\oplus}=10^5 \rho_{\rm DM}^{\rm G}$
and $\beta_\oplus c=20\,$km/s
and as long as $m_\phi\gtrsim 10^{-13}\,$eV the number of patches in the halo is large and we still assume that the field amplitude is stochastic.
Finally, we discuss an Earth halo phenomenological model, 
with 
$\rho_{\rm DM}^{\oplus}$, taking its maximally allowed value, that depends on the \SS{UDM-particle }
mass~\cite{BanerjeeComPhys2020}, and the UDM has a negligible \SS{velocity} dispersion velocity 
(see Supplemental Material [Supp. Mat.]).

{\em Experimental approach.}
\subsection*{Experimental approach}
\cut{Nuclear masses can be measured directly with several types of instruments, such as mass spectrometers and Penning traps. However, these currently have $10^{-11}$-level resolution are not 
suited for searching for fast variation.
In atoms the hyperfine energy depends on the proton mass.} 
\cut{Specifically, the hyperfine splitting of an atomic energy level is proportional to $g_I\alpha^{2}(m_{{\rm e}}/m_{{\rm p}})E_{\rm Ryd}$, where $g_I$ is the nuclear g-factor. Thus,}
Atomic clocks 
\cut{such as hydrogen masers, and laser-cooled Cs and Rb atomic clocks} 
can be used to search for oscillations of
\cut{ $\Lambda_{{\rm QCD}}$ and quark masses via} the proton mass and the nuclear g-factor. However, the accessible frequency range is $f_\phi\lesssim 1\,$Hz due to the operation mode of the clocks. 
\cut{We note that the index of refraction of a solid shows a dependence on the nuclear mass, but this dependence is weak \cite{Braxmaier2001,SavallePRL2020}.}

A recently suggested alternative approach in this context is spectroscopy of molecules \cite{Hanneke2021,AntypasQST2021}. Their transition frequencies contain contributions stemming  from changes in rotational and vibrational energy. Here we focus on the latter. 
The vibrational energy  $\hbar\omega_{{\rm vib}}$ of a diatomic molecule
containing two nuclei $N_{1},\,N_{2}$ scales approximately as $E_{{\rm Ryd}}\sqrt{m_{{\rm e}}/\mu}$,
where $\mu=m_{N_{1}}m_{N_{2}}/(m_{N_{1}}+m_{N_{2}})$ is the reduced nuclear mass. 
Thus, molecular transitions with a change of vibrational energy are sensitive to the nuclear mass. Furthermore, the electron-mass dependence is enhanced, beyond the scaling contained in $E_{\rm Ryd}\propto m_{\rm e}$. 

In a detection instrument based on spectroscopy, a reference quantum system having a resonance frequency $\nu^{(1)}$ is interrogated by the wave  of frequency $\nu^{(2)}$ emitted from an oscillator. 
$\nu^{(2)}$ is tuned to the proximity of $\nu^{(1)}$. In practice, the oscillator is often stabilized to another reference (atomic ensemble or cavity). Both frequencies may depend on more than one FC. The fluctuation spectrum of the frequency deviation $\Delta\nu(t)/\nu=[\nu^{(1)}(t)-\nu^{(2)}(t)]/\nu$ is measured.
The dependence of a frequency $\nu^{(i)}$ on a particular FC $g$ may be characterized by the fractional derivative $R^{(i)}_{g}=d\ln \nu^{(i)}/d\ln g$.
A hypothetical modulation $\delta g/g$ of a constant $g$ causes a modulation of the frequency deviation $\delta\nu/\nu=(R_{g}^{(1)}-R_{g}^{(2)})\delta g/g$. 
One key parameter of a given experiment is, therefore, the differential sensitivity $\Delta R_g=R_{g}^{(1)}-R_{g}^{(2)}$, determined by the choice of reference and oscillator. 

\subsection*{Apparatus and Operation} 
\begin{figure}[H]
\includegraphics[width=10cm]{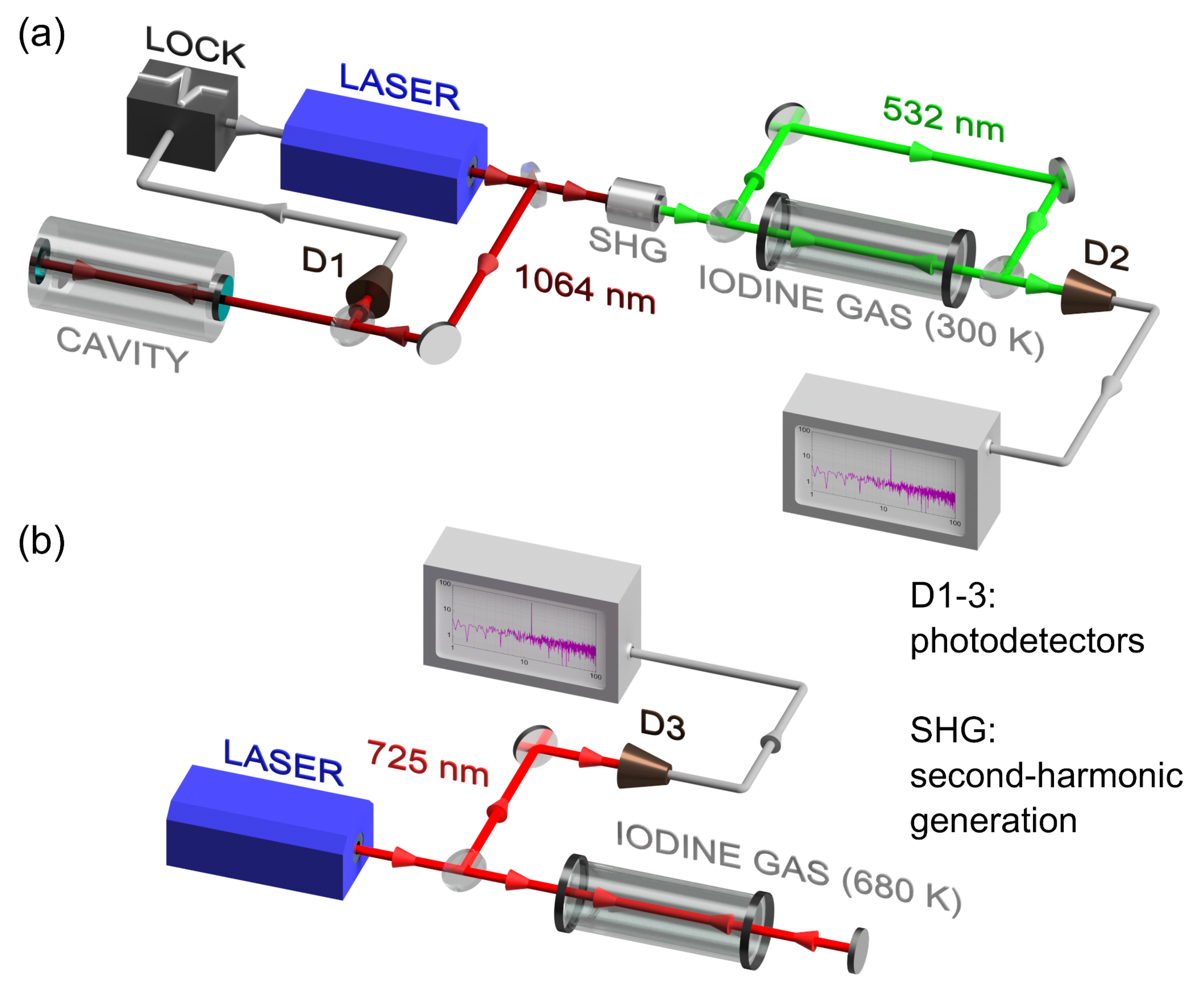}
\caption{\small{The two molecular iodine experiments to search for FC oscillations. (a) Experiment A; (b) Experiment B. 
} }
\label{fig:Both3DSetups}
\end{figure}

In our experiments, we use an electronic transition of molecular iodine (I\textsubscript{2}) between the  ground electronic state X and the  excited electronic state B \cite{Gerstenkorn1985}. The concepts of the
experiments are shown in Fig.\,\ref{fig:Both3DSetups}. Details are presented in the Supp. Mat.
\cut{
The transition frequency can be approximated as 
$\nu^{(1)}=\nu_{0}+\nu_{{\rm vib,B}}-\nu_{{\rm vib,X}}$, where  $h\nu_{0}\simeq(hc_{})15769\,{\rm cm}^{-1}$ is the difference in the electronic binding energies of the two states and $h\nu_{{\rm vib}}$ is the vibrational energy. 
Rotational energy contributions can be neglected, due to the large mass of iodine. We approximate the vibrational energies as $h\nu_{{\rm vib,X}}=h\omega_{{\rm vib,X}}(\upsilon+1/2)$, $h\nu_{{\rm vib,B}}=(\upsilon'+1/2)h\omega_{{\rm vib,B}}$, with the vibrational constants $\omega_{{\rm vib,X}}=c_{}\,214.5\,{\rm cm}^{-1}$,
$\omega_{{\rm vib,B}}=c_{}\,125.7\,{\rm cm}^{-1}$. The vibrational quantum numbers in the states X and B are $\upsilon$ and $\upsilon'$, respectively.
It is reasonable to assume that the electronic energy difference arises mostly from non-relativistic dynamics. Since both electronic and vibrational energies are proportional
to the Rydberg energy, we have $R_{\alpha}^{(1)}=2$. The sensitivity to the electron mass is 
$R_{{\rm e}}^{(1)}\simeq1+(\upsilon'\omega_{{\rm vib,B}}-\upsilon\omega_{{\rm vib,X}})/2\nu^{(1)}$
and to the nuclear mass 
$R_{\rm N}^{(1)}\simeq-(\upsilon'\omega_{{\rm vib,B}}-\upsilon\omega_{{\rm vib,X}})/2\nu^{(1)}$.
Since the vibrational energy contribution is only a small fraction of the total transition energy, $R_{N}^{(1)}$ is small. In the future, the value can be increased by using pure vibrational transitions \cite{AntypasQST2021}.
}
We have performed two experiments, A and B. In the apparatus A the reference is the well-known transition R(56)32-0 at 532~nm ($\upsilon=0$,\,$\upsilon'=32$), with sensitivity $R_{\rm N}^{(1,{\rm A})}\simeq-0.06$. In apparatus B, the transition is R(122)2-10 at 725\,nm ($\upsilon=10$,\,$\upsilon'=2$) \cite{RakowskyAPB1989}, with $R_{\rm N}^{(1,{\rm B})}\simeq0.07$.

In both experiments, the interrogating oscillator is a laser (frequency $\nu^{(2)}$). Apparatus A uses a monolithic Nd:YAG laser whose frequency is actively stabilized to a reference cavity. 
\cut{made of ultralow thermal expansion glass (ULE) with the optical cavity mode 
in vacuum. }
\cut{The detected 
frequency range covers 10\,Hz to 100\,kHz. The lower end of this range covers frequencies smaller than the bandwidth of the frequency lock of the laser to the cavity, $f_1^{(\rm A)}=3$\,kHz. For $f<f_1^{(\rm A)}$, the frequency $\nu^{(2,{\rm A})}$ of the wave sent to the experiment is determined by the length of the ULE cavity. As widely discussed, for a cavity}
For this configuration, and the considered frequency range, $R_{\alpha}^{({\rm 2})}=1$,
$R_{\rm e}^{({\rm 2})}=1$ (see Supp. Mat.).  
\cut{\cite{Turneaure1983} For $f>f_1^{(\rm A)}$ the laser resonator is the element determining the frequency fluctuations. Although the resonator is monolithic and the material has a refractive index larger than unity, to a good approximation, the same expressions for $R_{\alpha}, R_{\rm e}$ hold. 
We  neglect the effect of mechanical resonances \cite{SavallePRL2020}.
}
In apparatus B, the laser is a Ti:Sapphire laser and 
frequencies in the range 100\,kHz--100\,MHz are considered. As this range is above the acoustic cutoff frequency of the laser $f_2^{({\rm B})}\simeq50\,$kHz \cite{AntypasPRL2019},  the frequency $\nu^{(2,{\rm B})}$ is essentially independent of the FCs \cite{{Antypas:2020rtg}}. 

Summarizing, experiments A and B provide sensitivity to $\alpha$, $m_{\rm e}$, and $m_{\rm N}$. For experiment A,  
$\Delta R_{\alpha}^{({\rm A})}\simeq2-1=1$, 
$\Delta R_{\rm e}^{({\rm A})}=
(1-R_{\rm N}^{(1,{\rm A})})-1\simeq +0.06$,
$\Delta R_{\rm N}^{({\rm A})}=
R_{\rm N}^{(1,{\rm A})}-0\simeq-0.06$,
while for experiment B, 
$\Delta R_{\alpha}^{({\rm B})}\simeq2$, 
$\Delta R_{\rm e}^{({\rm B})}=
(1-R_{\rm N}^{(1,{\rm B})})-0\simeq +0.93$ 
$\Delta R_{\rm N}^{({\rm B})}=
R_{\rm N}^{(1,{\rm B})}-0
\simeq 0.07$. 
In both experiments, the instantaneous frequency deviation $\Delta\nu$ is converted into a voltage signal  $V^{(k)}(t)=D^{(k)} \Delta\nu^{(k)}(t)$,  with the discriminators $D^{(k)}$ being system parameters, and $k={\rm A,B}$. 
 This allows us to obtain the spectrum of the fractional frequency variation $\delta \nu ^{(k)}/\nu^{(k)}$.  The time-varying FC ($\alpha$, $m_{\rm{ e}}$ and $m_{\rm {N}}$) contribute to the variation according to: 
%
\begin{equation}
\frac{ \delta \nu^{(k)}}{\nu^{(k)}}= 
	\Delta R_{\alpha}^{({k})}\frac{\delta \alpha}{\alpha} + \Delta R_{\rm{e}}^{({ k})}\frac{\delta m_{\rm e}}{m_{\rm e}} +
	\Delta R_{\rm{N}}^{({k})}\frac{\delta m_{\rm N}}{m_{\rm N}}\ .
\label{eq:molfrequencyvariation}
\end{equation}
\noindent 
The Eq.~\,\eqref{eq:molfrequencyvariation} is used to probe oscillations of the FC.  
The cavity vs. molecular transition comparison can be used to constrain a combination of several 
DM - SM coupling constants.
\cut{If only one coupling $d_g$ is nonzero, one would obtain its value from
$\Delta R_g^{(k)} d_g=\phi(t)^{-1}\delta\nu(t)^{(k)}/\nu^{(k)}$, with $\phi(t)$ displayed in full detail in Supp. Mat..}

\subsection*{Search for oscillating fundamental constants} 
\begin{figure}[H]
\begin{centering}
    \begin{subfloat}
    { 
            \includegraphics[width=14cm]{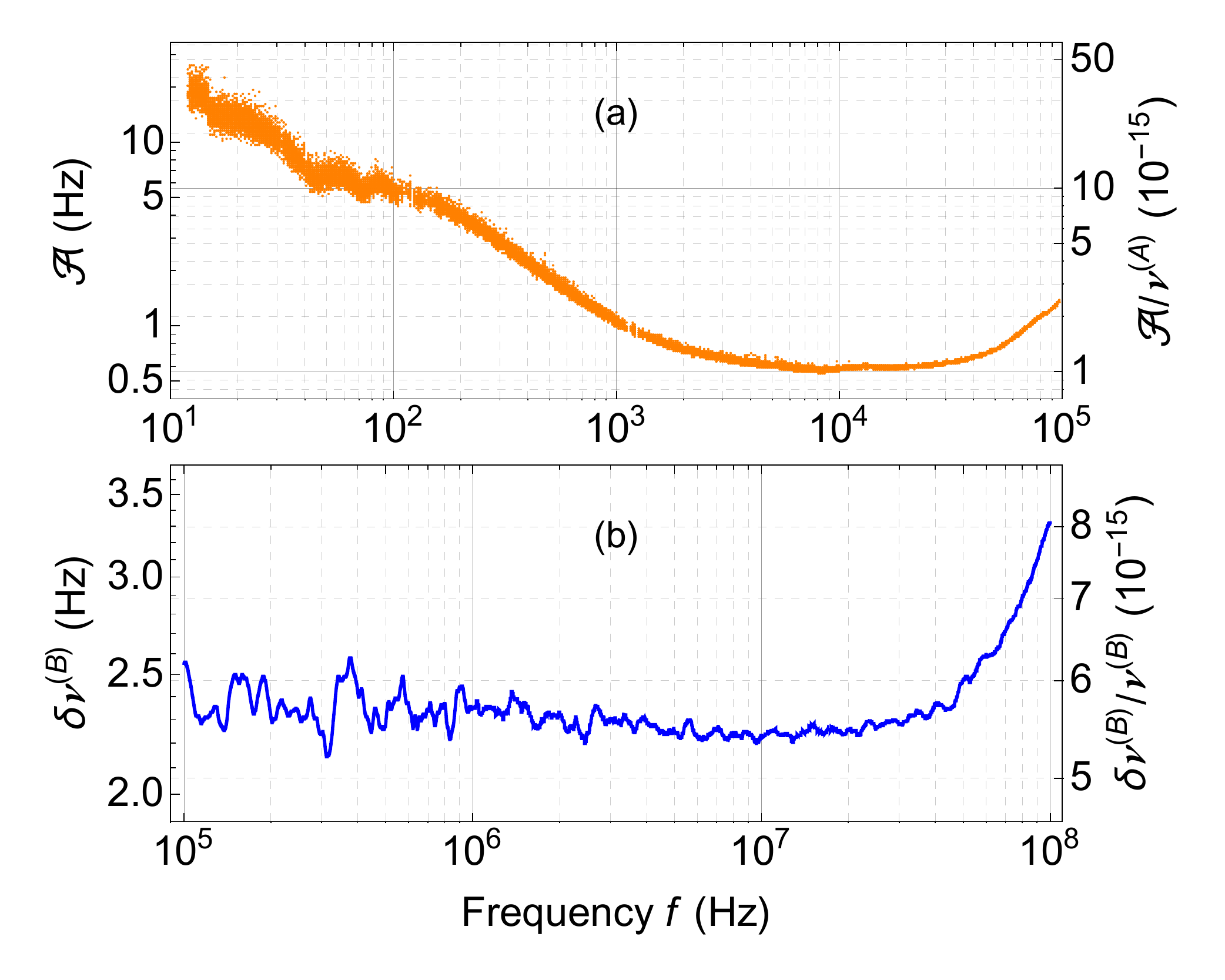}
        \label{fig:df_f_A}
    }
    \end{subfloat}\\


\end{centering}
\caption{
  \small{
  {\bf (a)}: Experiment A. The   
  spectral amplitude ${\cal A}=$ $\sqrt{{\rm PSD_F}/T}$ of the scaled discriminator signal $\Delta\nu^{(\rm A)}(t)=V^{(\rm A)}(t)/D^{(\rm A)}$. 
  ${\rm PSD_F}$ is the optimally filtered power spectral density (PSD), with a filter chosen appropriately for signals having the same linewidth $f/Q_0$, $Q_0\simeq1\times10^6$, as standard galatic halo UDM. The width of the orange band corresponds to the mean of ${\cal A}\pm\sigma(f)$.
{\bf (b)}: Experiment B. The
    bound (95\% confidence level) on fluctuations of the signal $\Delta\nu^{(B)}(t)=V^{(\rm B)}(t)/D^{(\rm B)}$. 
 }
}
\label{fig:Both_df_f}
\end{figure}

In the experiments, the lasers are tuned to the respective iodine transitions and the signals $V^{(k)}(t)$ are recorded.  
\cut{At present, we achieved uninterrupted operation for up to 50\,h for experiment\,A and 60\,h for experiment\,B.} 
In experiment\,A, the voltage $V(t)$ was recorded continuously with a 16-bit data acquisition (DAQ) system and a sampling rate of $250\,$kSa/s.
Here, $D\simeq1\,$V/MHz. We analyzed a set of $N=2^{34}$ \cut{consecutive} samples spanning $T\simeq19\,$h.
From the data $V_i=V^{\rm(A)}(t_i)$ the normalized periodogram $P_k=|\tilde{V_k}|^2/N^2$ was calculated. This is the same as the discretized power spectral density (PSD), multiplied by 
\cut{the frequency-bin width} 
$1/T$.  
Various peaks in the periodogram were investigated and identified as being of technical origin, 
in part 
by shifting the interrogation-laser frequency away from the resonance and repeating the measurement. This left no obvious candidate UDM signals in the spectrum. \SS{A number of} 
frequency intervals exhibiting spectral peaks of technical origin are listed in the Supp. Mat. We do not give limits for these excluded intervals, \SS{that have widths of 5\,Hz or smaller.} 
From the periodogram,  
 the upper limit of the coupling parameters $d_{g}$ 
was determined using  the analysis 
of Ref.\,\cite{Derevianko2018}.
\cut{, and report the results further below}
The spectral amplitude of the recorded signal is shown \cut{One suitable display of the spectrum of the recorded data is 
a  convolution of $P_k$ with  a line-shape function that emulates the UDM spectrum. The UDM spectrum 
is model-dependent so for concreteness we consider the standard galactic halo model \cite{Brubaker2017, Gramolin2021}, and for simplicity approximate the UDM spectrum by a Lorentzian with quality factor $Q=1\times10^{6}$. 
The convolution procedure results in an optimally-filtered periodogram (OFP); it enhances the detectability of UDM-like signals. The square root of the OFP, multiplied with $D$, (ROFP) is 
shown in a simplified manner} 
in Fig.\,\ref{fig:df_f_A}. 
For interpretation see Supp. Mat.
\cut{The trend of the mean of ROFP as a function of Fourier frequency $f$ is due to various noise sources, in particular the increase for $f>10^4\,$Hz is due to laser amplitude noise.}

In experiment B, the voltage $V(t)$ was measured with the laser frequency tuned either on the slope of the I$_2$ resonance, or off-resonance, alternating between these UDM-sensitive and insensitive acquisition modes 
to account for spurious signals due to sources other than UDM. A 12-bit DAQ system sampled $V(t)$ in successive $T=0.1$\,s-long intervals at a rate of 250\,MSa/s and the corresponding periodograms were computed and continuously averaged. 
The  periodogram difference between the on- and off-resonance acquisition modes was also computed and averaged over a 60-hr-long run.  This spectrum will contain power in excess of statistical noise in the presence of FC oscillations, and it is subsequently investigated for UDM detection. A number of candidate peaks were identified with power in excess of a 95\% detection threshold, which was computed with consideration of the `look elsewhere' effect \cite{ScargleTAJ1982}.
All spurious signals were checked in auxiliary experiments and eventually attributed to technical noise. The post-inspected spectrum is used to obtain a constraint for $\delta\nu^{(\rm B)}/\nu^{(B)}$ that is shown in Fig.\,\ref{fig:Both_df_f}\,b. See Supp. Mat. for details.

\cut{{\emph{
Models and bounds--}} 
}
We analyse the experimental data in the framework of  three mentioned UDM models that differ in terms of the amplitude of the UDM field and its
\cut{ coherence time of the UDM field:
(1) the galactic halo model, the standard case addressed in many previous works \cite{Foster2018,Derevianko2018} with $\rho_{\rm DM}^{\oplus}=\rho_{\rm DM}^{\rm G}$;
(2) the Sun halo model (adiabatic contraction) \footnote{This model is motivated by~\cite{Anderson:2020rdk} and some ongoing work.}. The UDM linewidth is reduced compared to (1); this originates from a virial velocity equal to the escape velocity of the Earth in the solar system, $v_0^{\rm(S)}\simeq 20\,{\rm km/s}\,.$ 
}
\cut{(3) the Earth halo model. This may alternatively be called the "gravitational hydrogen atom model". The UDM field is monochromatic with infinite coherence time, following from the assumption that the Earth halo is infinitely stable, i.e. the virial velocity is zero.
The value of $\rho_{\rm DM}^{\rm E}$ is a function of UDM particle mass (see Fig.\,2a in supplementary information of Ref. \cite{BanerjeeComPhys2020}), and is enhanced compared to $\rho_{\rm DM}^{\rm G}$ by a factor increasing from $10^4$ at $f_{\phi}=100\,{\rm Hz}$ to $10^{19}$ at $f_{\phi}=3.4\,{\rm MHz}$. However, beyond $f_\phi\simeq15\,$MHz the ratio $\rho_{\rm DM}^{\rm(E)}/\rho_{\rm DM}$ drops below 1.
For these different scenarios the}
coherence time $\tau_{\rm coh} = 1/(\omega_\phi (v_{\rm vir}/c)^2)$: 
\begin{eqnarray}
\tau_{\rm coh} = 
\begin{cases} 
5.9\times 10^5 f_\phi^{-1},\ Q=1.1\times10^6,
\,\,{\rm  galactic\,halo}\\ 

7.1\times 10^7 f_\phi^{-1},\ Q=9.0\times10^7,
\,\, {\rm solar\, halo}\\

\infty,\,\, {\rm Earth\,halo}\,.
\end{cases}
\end{eqnarray}
\cut{Although the continuous acquisition time $T$ of our two experiments differs by orders of magnitude, the total measurement time of both experiments (tens of hours) exceeds the coherence times of the galactic halo and solar halo models for most of the considered range of $f_\phi$ values, so that the Rayleigh probability distribution of the amplitudes $\alpha_i$ is fully sampled. For the Earth halo model, there is but a single amplitude,  $\alpha_1=\sqrt{2}$ (See Supp. Mat. for details).}



\begin{figure}[H]
    \vskip 2mm
        \centering
        \includegraphics[width=14cm]{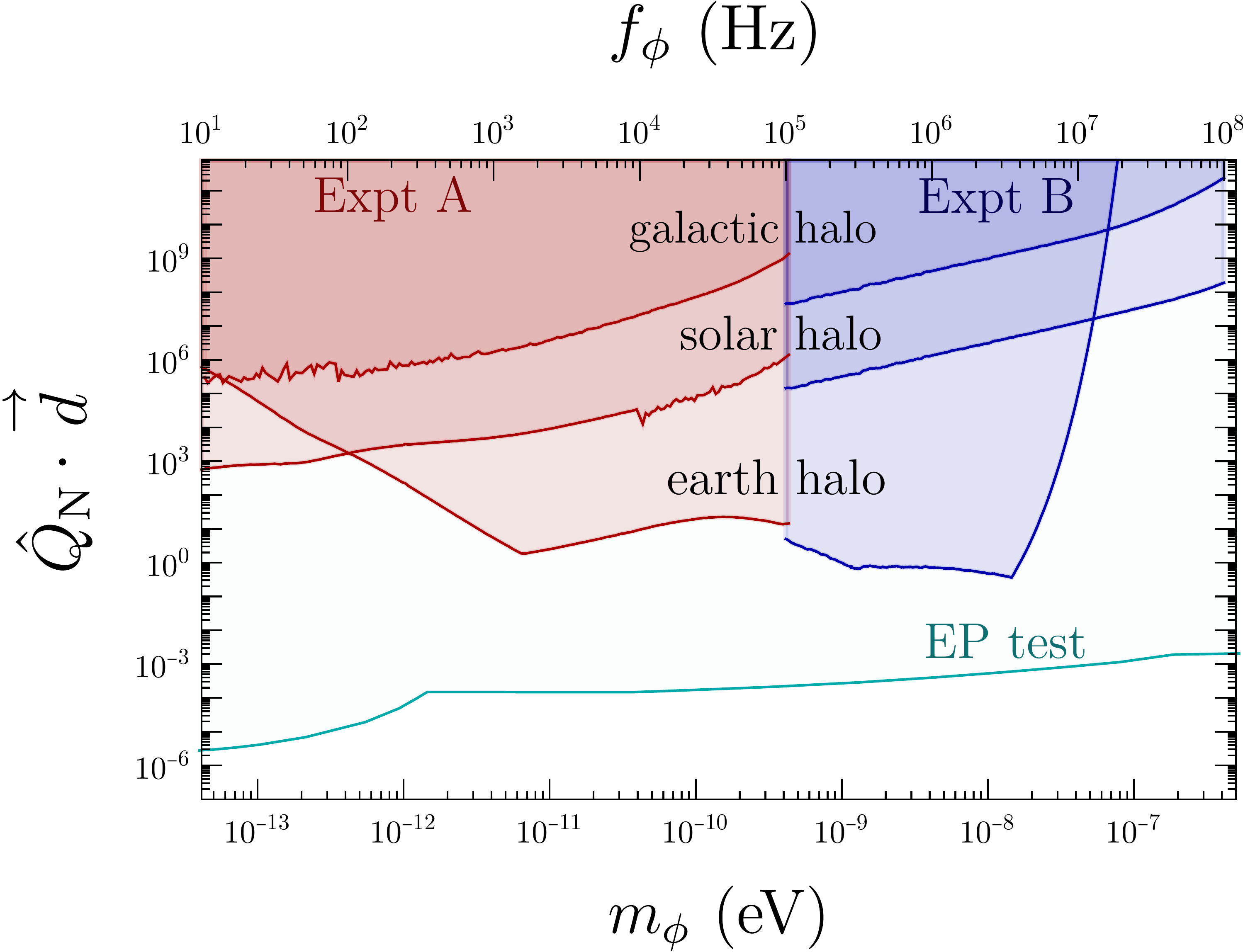}
    \label{fig:dtotal}
     
    \caption{\label{fig:bkg_DM}{\small 
    Exclusion plot of the combination of couplings to the QCD sector, $\hat Q_{\rm N}\cdot\vec d=0.999\, d_{g_{\rm s}}+0.092\,d_{\hat m}+3\times10^{-4}\,d_{\delta m}+0.047\,d_{m_{\rm s}}$. 
\cut{as a function of scalar field mass or Compton frequency. The red and the blue lines are the constraints coming from experiment A and experiment B respectively, while 
the region above the} 
Turquoise line: \cut{is excluded by various} fifth-force/EP-violation experiments 
\cut{searching for the existence of a light scalar particle}~\cite{PhysRevD.61.022001,Schlamminger:2007ht,Wagner:2012ui,Berge:2017ovy,HeesPRD2018}.
}}
\end{figure}

In order to derive bounds to the UDM couplings, we assume that only one of the constants $m_e$, $\alpha$, or $m_{\rm N}$ in Eq.\,\eqref{eq:molfrequencyvariation} oscillates and analyze the three cases separately. See Supp. Mat. for details.
In Fig.~\ref{fig:bkg_DM}, we present our constraints together with existing EP constraints (turquoise line) on the combined quark and gluons couplings $\hat Q_{\rm N}\cdot \vec d$ 
\cut{[see Eq.~\eqref{eq:varnuclearmasswithPhi}]}
as a function of the UDM mass. 
\cut{The red and the blue lines refer to experiment A and experiment B, respectively. The lines and the texts within depict the scenario with the standard galactic UDM halo, the solar halo and the Earth halo model respectively. }
\cut{As mentioned above, the presence of a light scalar also induces a Yukawa force and a potential violation of the EP that is subject to strict experimental bounds, see
the turquoise line.} 
\cut{represents bounds from various EP tests on our scenario~\cite{PhysRevD.61.022001,Schlamminger:2007ht,Wagner:2012ui,Berge:2017ovy,HeesPRD2018}. 
}

\begin{figure}[H]
  \includegraphics[width=14cm]{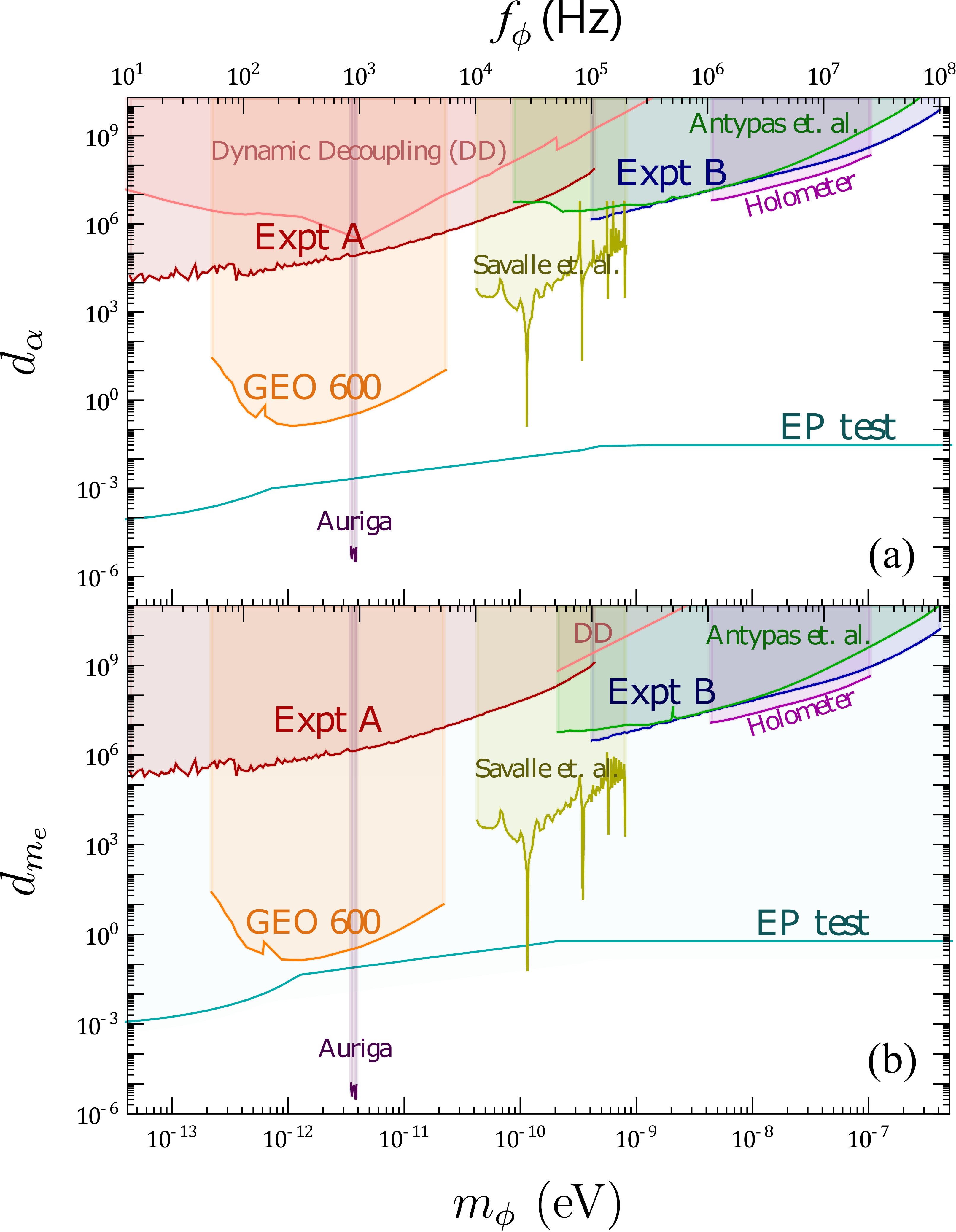}

    \caption{\small{Exclusion plot of the coupling to (a) $\alpha$ and (b) $m_{\rm e}$. 
    \cut{as a function of the scalar field mass. The red and the blue shaded regions are the constraints from experiment A and experiment B, respectively.} 
Existing constraints: \cut{are denoted by the}  shaded regions in orange~\cite{VermeulenArxiv2021}, yellow~\cite{Savalle:2020vgz}, pink~\cite{AharonyPRD2021}, green~\cite{Antypas:2020rtg}, Magenta~\cite{Aiello:2021wlp}, and purple~\cite{PhysRevLett.118.021302}. \cut{Constraints coming from fifth force/EP tests are denoted by the turquoise shaded region}
Fifth-force/EP tests: turquoise~\cite{PhysRevD.61.022001,Schlamminger:2007ht,Wagner:2012ui,Berge:2017ovy, HeesPRD2018}.
}}
\label{fig:dmedalpha}
\end{figure}

Constraints on the variation of $\alpha$ and $m_e$ and $\alpha$ are presented in Fig.~\ref{fig:dmedalpha}a and Fig.~\ref{fig:dmedalpha}b, respectively, alongside previous constraints. 
We  show only the strongest existing constraints on the relevant parameter space. 
In principle, astrophysical bounds on our scenario could also apply, however, these are typically weaker than those discussed here and are less robust
 (see~\cite{PhysRevD.104.015012,PhysRevD.102.075015,PhysRevD.101.123025} for recent discussions).
For the sake of 
clarity, these constraints are presented for the standard galactic UDM halo only.
Our results cover the previously unexplored bands  $10-50$\,Hz and  $5-10$\,kHz, and improve on existing bounds in the range $5-100$\,MHz. 

\begin{figure}[H]
    
        \includegraphics[width=14cm]{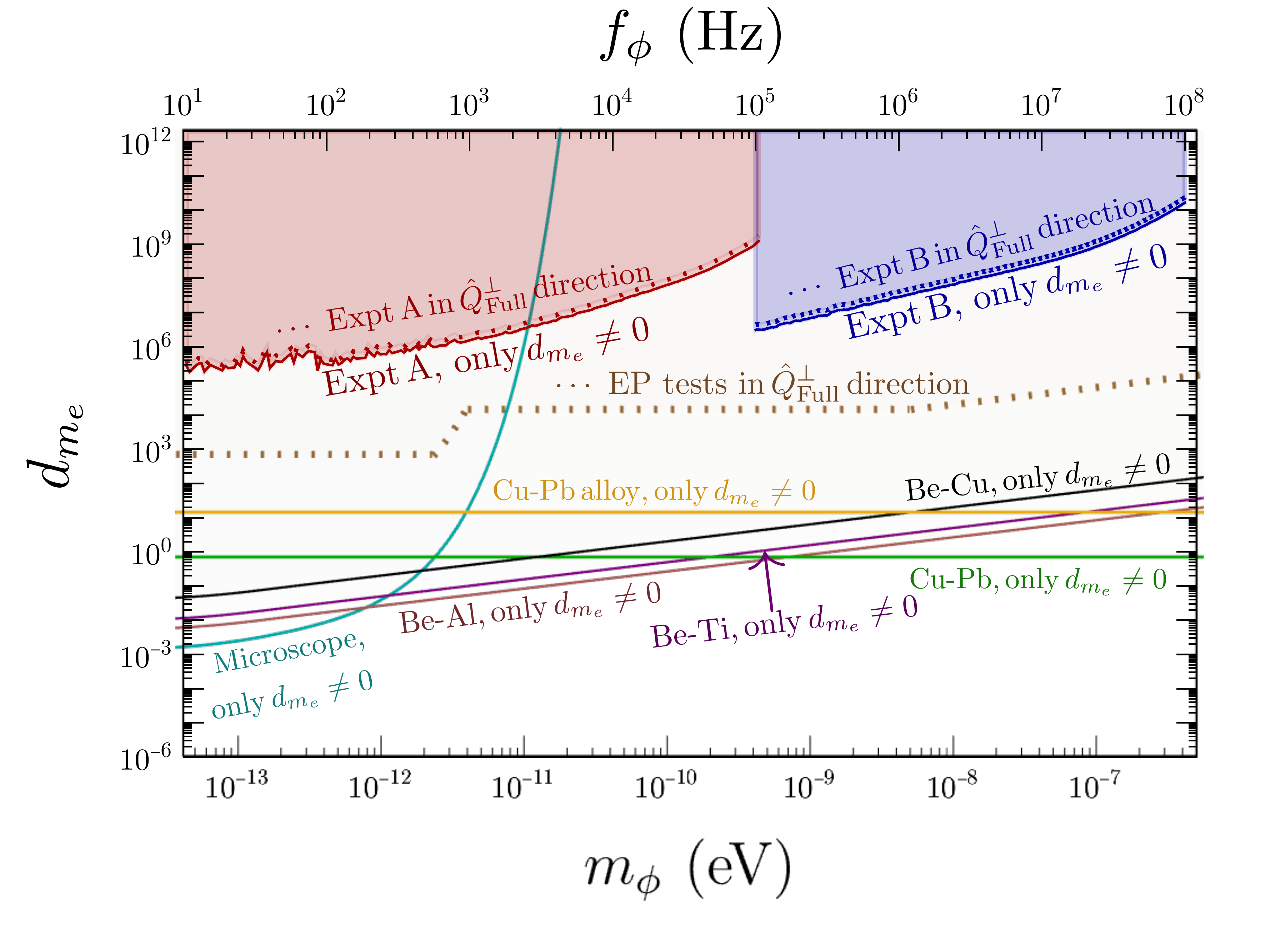}
    
    \caption{\small{
    \cut{{\bf (a):}
Exclusion plot for the coupling constant $d_{g_{\rm s}}$. Solid lines assume a model where only $d_{g_{\rm s}}\neq0$; dotted lines give the bounds for the linear combination of couplings $\hat Q_{\rm QCD}^\perp\cdot\vec d_{\rm QCD}$, chosen to be orthogonal to the Cu-Pb~\cite{PhysRevD.61.022001}, and Be-Al~\cite{Wagner:2012ui} EP tests. 
The red and the blue colors depict bounds from our experiment A and B respectively. 
Be-Ti~\cite{Schlamminger:2007ht}, and the MICROSCOPE experiment~\cite{Berge:2017ovy} provide best EP violation bounds, and are shown in magenta and turquoise dotted lines respectively. The Be-Cu~\cite{Su:1994gu} experiment, shown in black color provides weaker bounds.
}
Exclusion plot for $d_{m_e}$; the solid lines assume a model where only $d_{m_e}\neq0$. 
The dotted lines depict the bounds for a model defined by a vector of sensitivities, $\hat Q_{\rm Full}^\perp(m_\phi)$, that is orthogonal to the sensitivities of four leading EP test experiments (see Supp. Mat. for details).
The bounds from our experiments are shown in red and blue, whereas bounds from other published experiments, shown in Fig.~\ref{fig:dmedalpha}, are not shown again here, for simplicity. 
The bound from the fifth-best EP test experiment in a given mass range, projected onto the $\hat Q^\perp_{\rm Full}(m_\phi)$ direction, and further on the $d_{m_{\rm e}}$ direction, is shown as dotted brown line. 
}}
\label{fig:dmedirection}
\end{figure}

More generally, our 
 experiments are sensitive to the following linear combinations of the full set of couplings, defined in Eq.~\eqref{eq:varnuclearmasswithPhi},\eqref{eq:molfrequencyvariation}, that can be written as 
$\vec Q\cdot \vec d\,=|\vec Q\,|\hat Q\cdot\vec d$ 
with $|\vec Q^{\rm A}|=1\,$, $|\vec Q^{\rm B}| = 2.21\,$, and
\begin{eqnarray}
\hat Q^{\rm A}\simeq&&\,\,(1.0,0.06,0.05,0.005,1.5\times10^{-5},0.0024)\ ,\nonumber \\ 
\hat Q^{\rm B}\simeq&&\,\,(0.90,0.42,-0.027,-0.0025,-8\times10^{-6},-0.0013)\,.\nonumber
\end{eqnarray}
We now discuss the complementarity between direct UDM searches and the bounds arising from EP tests.
\cut{First, consider QCD-philic models, where the UDM only couples to the QCD sector, i.e. we assume that only $d_{{g_{\rm s}},\hat m,\delta m,m_{\rm s}}\neq0$. For the present experiments, the corresponding sensitivities are $\vec Q^{\rm A,B}$, projected onto this 4-dimensional coupling parameter space. 
As we have already discussed, the EP bounds will be sensitive to the differences in the dilatonic charges of test bodies, characterised by }
\cut{$(\Delta \hat Q)^{\rm Exp} \cdot \vec d\,$ (see Supp. Mat. for details).
Examining the dependence of the EP test experiments one finds that the dependence of 
$(\Delta \hat Q)^{\rm Exp}$ on $d_{g_{\rm s}}$ and $d_{m_{\rm s}}$ are equal which leaves us with effective vector space of three independent couplings, $\vec d_{\rm QCD} = (d_{g_{\rm s}},d_{\hat m},d_{\delta m})\,$.
Thus, one can choose a direction in this coupling vector space, $\hat Q^\perp_{\rm QCD}=\left(0.15,-0.03,0.99\right)$, that is orthogonal to the two most sensitive EP tests, Cu-Pb~\cite{SmithPRD1999} and Be-Al~\cite{Wagner:2012ui} (see Supp. Mat. for details). 
The resulting bound, projected for simplicity onto the $d_{g_{\rm s}}$ direction, is shown as the dotted lines of Fig.~\ref{fig:dgdirection}. 
We find that, while the bound from our direct UDM experiments weakens by a factor  $\approx7$ compared to the case when only $d_{g_{\rm s}}\ne0$, the corresponding EP bound (Be-Ti, dotted) weakens by more than two orders of magnitude in the large-mass region, $f_\phi\simeq 100\,{\rm MHz}$.}
\cut{Second, consider models where the}
Since UDM is allowed to couple to all fields, the relevant parameter space is of dimension five. 
One can 
find a direction $\hat Q^\perp_{\rm Full}(m_\phi)$ in this space that is orthogonal to the best four EP-test bounds for a given mass.  
For example, in the mass range of $2\times 10^{-12}\lesssim m_{\phi}/{\rm eV}\lesssim 5\times 10^{-9}$, these are the 
Be-Al~\cite{Wagner:2012ui}, Be-Ti~\cite{Schlamminger:2007ht},  Cu-Pb~\cite{SmithPRD1999} and Be-Cu~\cite{Su:1994gu} experiments. 
and we find $\hat Q^\perp_{\rm Full}(m_\phi)\simeq \big(
0.003\,,\, -0.987\,,\, 0.002\,,\,-0.001\,,\,-0.162\,\big)\,.$
For masses above $5\times 10^{-9}\,$eV, $\hat Q^\perp_{\rm Full}$ is perpendicular to the Be-Al~\cite{Wagner:2012ui}, Be-Ti~\cite{Schlamminger:2007ht},  Cu-Pb~\cite{SmithPRD1999} and Cu-Pb-alloy~\cite{Nelson:1990uk} \showgray{tests} \SS{ sensitivity vectors,} with corresponding $\hat Q^\perp_{\rm Full}$.
Models of light scalar UDM with coupling direction defined according to $\hat Q^\perp_{\rm Full}(m_\phi)\cdot \vec d$ would not be constrained by these four leading EP bounds. Note that throughout the whole the mass range $\hat Q^\perp_{\rm Full}(m_\phi)$ has a substantial overlap with the $d_{m_e}$ direction (the second entry). 
Thus, experiments that are particularly sensitive to time-variation of $m_{\rm e}$, such as the ones being discussed here, test a sector of coupling space that the first four-best EP experiments are insensitive to.
In Fig.~\ref{fig:dmedirection}, \showgray{for clarity,} we present our bounds on $\hat Q^\perp_{\rm Full}\cdot \vec d$, projected \SS{(for clarity)} in the $d_{m_e}$ direction, as dotted red and blue lines. 
The fifth-best EP bound projected onto $\hat Q^\perp_{\rm Full}(m_\phi)$ 
and further on $d_{m_{\rm e}}$, is shown by a brown dotted line. Note that we could only calculate the projection of $\hat Q^\perp_{\rm Full}$ into the remaining 5th-best EP bound to an accuracy of 1:10$^3$ \showgray{as we were limited by the information given by experiments done some 3 decades ago.} \SS{ due to the limited precision of the published test mass composition data.} 
We find that in this sector of coupling space the bounds related to our direct UDM experiments are \SS{only} two to three orders weaker than the bounds from the EP tests. 

\subsection*{Conclusion}
Our molecular-spectroscopy experiments have resulted in the first bounds on the coupling of an oscillating UDM field to the gluon and quark fields, in a broad frequency range  
that
spans seven decades (10\,Hz--100\,MHz). Within this range, improvements on previous limits for the coupling to the electromagnetic field and the electron field were also obtained within a small frequency window. 
Our experiments are not entirely free of technical noises. A new generation of similar experiments, with minimization of all noise sources, long acquisition times, high sampling rates, and, possibly, multiple setups enabling reduction of the noise level by statistical averaging, could further improve the present limits by several orders of magnitude. Furthermore, we have argued that there is a special class of dark matter couplings where the bounds from equivalence principle tests are significantly \showgray{ameliorated.}\SS{ less stringent than expected.}
Consequently, \showgray{it opens up the possibility that}  in the near future experiments of the kind described here \showgray{will} \SS{may }be able to probe \showgray{uncharted territories of} this class of UDM models \SS{with sensitivity competitive to EP tests.}



\begin{acknowledgments}
\emph{Acknowledgments.} We thank D. Kanta for help with the project, and D. Iwaschko and R. Gusek for electronics development. 
A.B. thanks the Galileo Galilei Institute for Theoretical Physics for the hospitality and the INFN for partial support during the completion of this work.
This work was supported by the Cluster of Excellence ``Precision Physics, Fundamental Interactions, and Structure of Matter'' (PRISMA+ EXC 2118/1) funded by the German Research Foundation (DFG) within the German Excellence Strategy (Project ID 39083149), by the European Research Council (ERC) under the European Union Horizon 2020 research and innovation program (project Dark-OST, grant agreement No 695405, project YbFUN, grant agreement No 947696), by the DFG Reinhart Koselleck project. The work of A.B. was supported by the Azrieli Foundation.
The work of G.P. is supported by grants from BSF-NSF (No. 2019760), Friedrich
Wilhelm Bessel research award, GIF, the ISF (grant No.\,718/18), Minerva, SABRA - Yeda-Sela - WRC Program,
the Estate of Emile Mimran, and The Maurice and Vivienne Wohl Endowment.
\end{acknowledgments}

%
\bibliography{Manuscript_v4b.bib}

\ifarXiv
    
    

\section*{Supplementary material}

# Supplemental Material:
# Search for oscillations of fundamental constants using molecular spectroscopy

R. Oswald, A. Nevsky, V. Vogt, S. Schiller*

*Institut für Experimentalphysik, Heinrich-Heine-Universität Düsseldorf, 40225 Düsseldorf, Germany*

N. L. Figueroa, K. Zhang, O. Tretiak, D. Antypas

*Johannes Gutenberg-Universität Mainz, 55128 Mainz, Germany and*
*Helmholtz-Institut, GSI Helmholtzzentrum für Schwerionenforschung, 55128 Mainz, Germany*

D. Budker

*Johannes Gutenberg-Universität Mainz, 55128 Mainz, Germany*
*Helmholtz-Institut, GSI Helmholtzzentrum für Schwerionenforschung, 55128 Mainz, Germany and*
*Department of Physics, University of California, Berkeley, California 94720, USA*

A. Banerjee, G. Perez

*Department of Particle Physics and Astrophysics,*
*Weizmann Institute of Science, Rehovot, Israel 7610001*

* step.schiller@hhu.de, kezhang@uni-mainz.de

# I. EXPERIMENTS

## A. Introduction

A molecular transition frequency can be approximated as $\nu^{(1)} = \nu_0 + \nu_{vib,B} - \nu_{vib,X}$, where $h\nu_0 \simeq (hc)15769\,\mathrm{cm}^{-1}$ is the difference in the electronic binding energies of the two states and $h\nu_{vib}$ is the vibrational energy. Rotational energy contributions can be neglected, due to the large mass of iodine. We approximate the vibrational energies as $h\nu_{vib,X} = h\omega_{vib,X}(v + 1/2)$, $h\nu_{vib,B} = (v' + 1/2)h\omega_{vib,B}$, with the vibrational constants $\omega_{vib,X} = c\,214.5\,\mathrm{cm}^{-1}$, $\omega_{vib,B} = c\,125.7\,\mathrm{cm}^{-1}$. The vibrational quantum numbers in the states X and B are $v$ and $v'$, respectively. It is reasonable to assume that the electronic energy difference arises mostly from non-relativistic dynamics. Since both electronic and vibrational energies are proportional to the Rydberg energy, we have $R_\alpha^{(1)} = 2$. The sensitivity to the electron mass is $R_e^{(1)} \simeq 1 + (v'\omega_{vib,B} - v\omega_{vib,X})/2\nu^{(1)}$ and to the nuclear mass $R_N^{(1)} \simeq -(v'\omega_{vib,B} - v\omega_{vib,X})/2\nu^{(1)}$. Since the vibrational energy contribution is only a small fraction of the total transition energy, $R_N^{(1)}$ is small. In the future, the value can be increased by using pure vibrational transitions [1]. Experiments A and B employ Doppler-free and Doppler-broadened $I_2$ transitions, respectively. The observed Doppler-free transition in apparatus A has a width of a few MHz. This width determines the frequency range over which there is significant molecular response to FC oscillations. In apparatus B, the observed Doppler-broadened transition is of the order of 1 GHz. In addition to Doppler broadening, there is significant homogeneous broadening due to collisions (as discussed below), and the molecular response is essentially constant over the 100 MHz range probed for FC oscillations. Apparatus A, being equipped with a low-frequency-noise laser source, is better suited for probing low frequencies up to the molecular transition's observed linewidth (of order MHz). Experiment B offers a broad detection range set by the transition's pressure broadening (hundreds of MHz). Important features of the experimental setups are the power spectral noise densities of the fractional frequency fluctuations of the molecular reference, $S_y^{(1)}(f)$, and of the interrogating laser wave, $S_y^{(2)}(f)$. These contain contributions of technical or of fundamental origin. In order to obtain small values of $S_y^{(1)}$ it is advantageous to employ references containing a large number of particles, here a substantial gas volume.

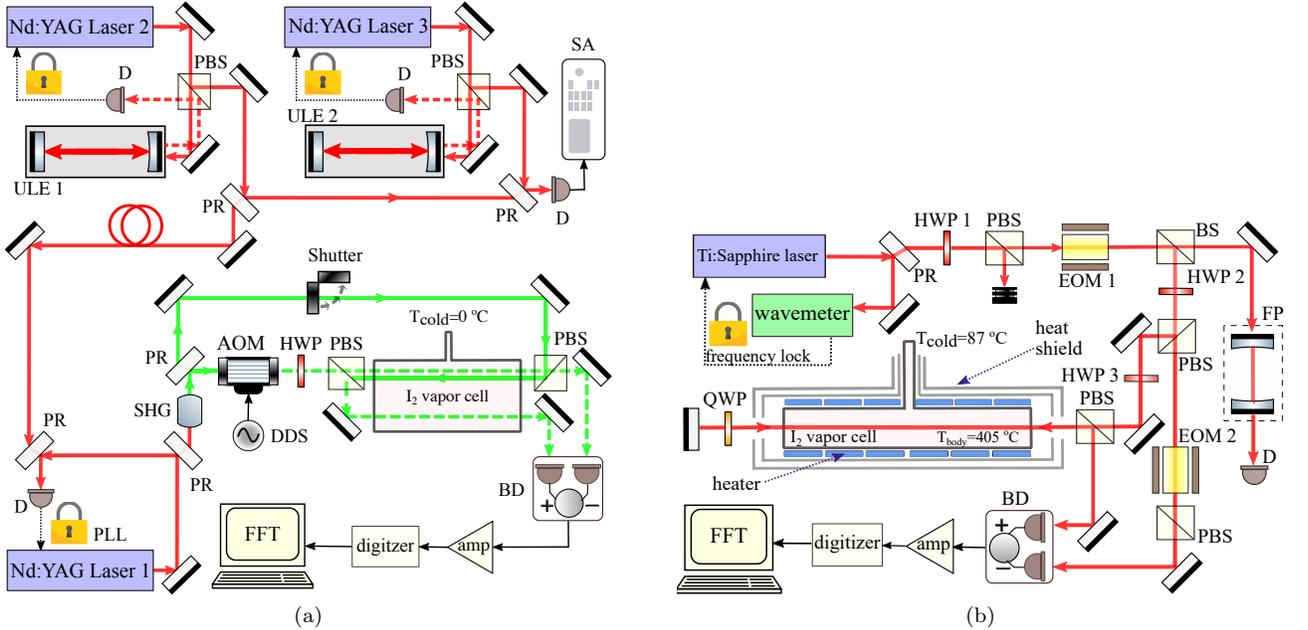

Figure 1. Experimental setups. (a) Setup for the experiment A. D: photodetector, AOM: acousto-optic modulator; SHG: second-harmonic generation; SA: spectrum analyzer; DDS: direct digital synthesizer; PBS: polarizing beam splitter; ULE: ultra-low expansion glass. BD: balanced photodetector; HWP: half-wave plate; PLL: phase-locked loop. (b) Setup for the experiment B. PR: Partial reflector; PBS: Polarizing beam splitter; BS: beam splitter, HWP: half-wave plate; QWP: quarter-wave plate; EOM: Electro-optic modulator. D: Photodetector; BD: Balanced photodetector; FP: Fabry-Perot cavity. HWP 3 is mounted on a stepper motor to allow for active balancing of the detector output.

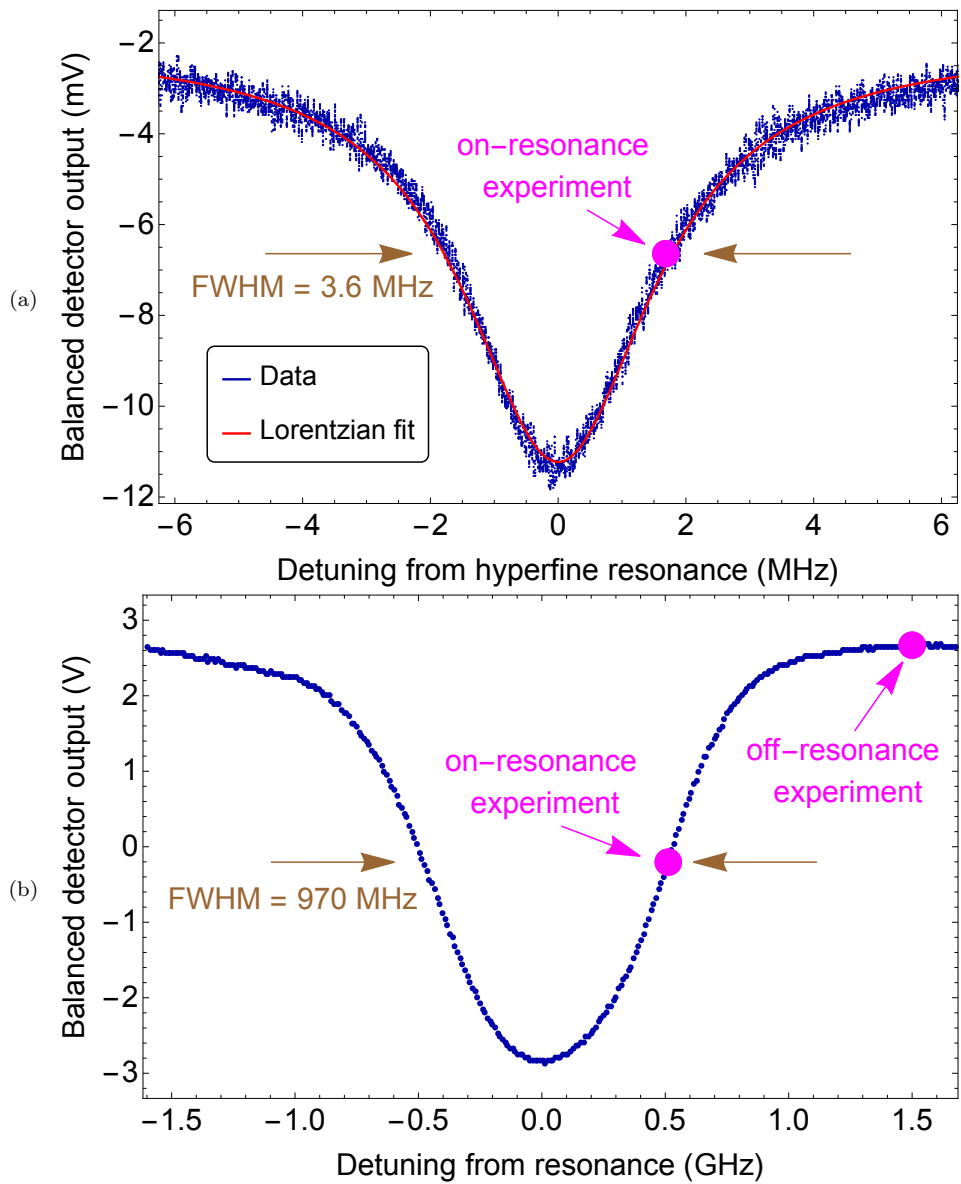

Figure 2. Spectra of iodine transitions employed in experiments A and B. (a): Doppler-free, pressure-broadened iodine transition at 532 nm in experiment A. FWHM: 3.6 MHz. Data was recorded with 1 MHz detection bandwidth and 10 ms scan time. For DM detection, the laser frequency is tuned to the operating point indicated by the magenta circle. The discriminator slope $D$ is found from the slope of the signal at this point and the electronics' amplification factor. (b): Doppler- and pressure-broadened $I_2$ transition at 725 nm in experiment B. The magenta-color circles indicate the regions where the laser frequency is tuned to take data with, or without sensitivity to FC oscillations. The spectrum was recorded with 250 Hz bandwidth and 2 s scan time.

### B. Experiment A

The oscillator interrogating the iodine gas is a Nd:YAG laser (laser 1). It is frequency-doubled to 532 nm by means of a fiber-coupled nonlinear conversion module. This wave has the frequency $\nu^{(2)}$. Laser 1 is phase-locked to a laser 2 that is frequency-stabilized to a 30 cm long ultrastable high-finesse ultra-low expansion glass (ULE) resonator.

The detected frequency range covers 10 Hz to 100 kHz. The lower end of this range covers frequencies smaller than the bandwidth of the frequency lock of the laser to the cavity, $f_1^{(A)} = 3$ kHz. For $f < f_1^{(A)}$, the frequency $\nu^{(2,A)}$ of the wave sent to the experiment is determined by the length of the ULE cavity. As widely discussed, for a cavity, $R_\alpha^{(2)} = 1$, $R_e^{(2)} = 1$ [2]. For $f > f_1^{(A)}$ the laser resonator is the element determining the frequency fluctuations. Although the resonator is monolithic and the material has a refractive index larger than unity, to a good approximation, the same

expressions for $R_\alpha, R_e$ hold. We neglect the effect of mechanical resonances [3].

Approximately 30 mW of laser radiation at 532 nm is sent to the spectroscopy setup. It is split into two beams (probe and saturation) using a 20%:80% beam splitter. These are sent from opposite directions into a 30 cm-long cell filled with iodine gas ($I_2$). The iodine saturation pressure in the cell is maintained at about 0.04 mbar by temperature stabilization of the cell's cold finger at about 0° C using a Peltier element. The probe laser beam is frequency-shifted by 50 MHz using an AOM to prevent interference effects with the saturating beam. The counter-propagating probe and saturation beams are overlapped inside the cell allowing nonlinear saturation spectroscopy on the hyperfine structure (hfs) components of iodine rovibronic absorption lines. Saturation absorption resonances are detected in the power of the transmitted probe beam by means of a low-noise photodetector D2.

To reduce the effect of laser intensity noise we implemented balanced detection of the probe wave and a reference wave. The latter is obtained by splitting off part of the probe beam using a half-wave plate (HWP) and a polarizing beam splitter (PBS) in front of the cell. This reference beam is sent through the cell colinearly with the spectroscopy probe beam, however not interacting with the saturation beam. The reference beam is detected with a photodetector D3, similar to D2.

The outputs of both photodetectors are subtracted using a precision differential buffer, and the signal is then further amplified by a factor of approximately $10^4$ using a low-noise preamplifier (Stanford Research Systems SRS 560). By adjusting the laser powers on the photodetectors by means of the HWP in front of the cell, the differential amplitude noise at the output of the amplifier can be reduced by about 40 dB over the bandwidth from DC to 1 MHz. The data were acquired using a 16-bit DAQ (digital acquisition) card (National Instruments USB-6343) referenced to a Maser.

In experiment A the spectroscopy was performed on the a1 hyperfine structure component of the R(56)32-0 electronic transition.

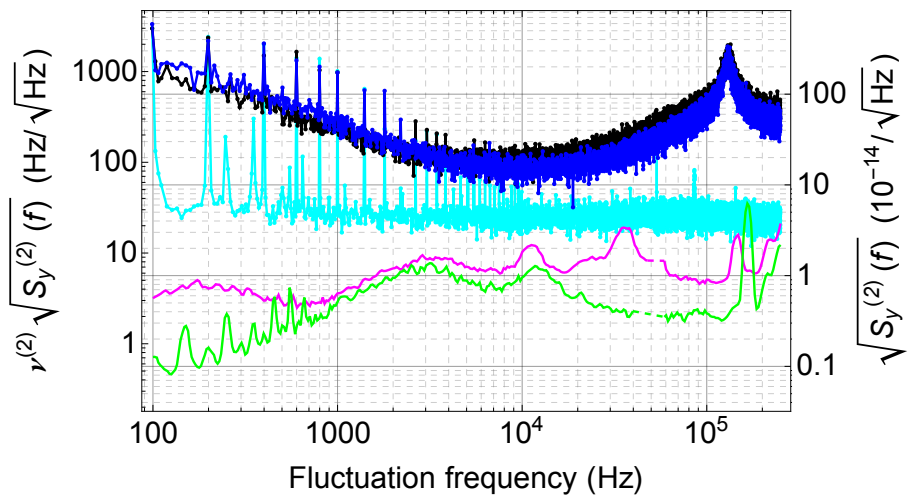

Figure 3.  total noise spectral density $\hat{S}_{\text{tot}}(f)$ for the detection of the relative frequency fluctuations between interrogating oscillator and iodine reference. The peak at 120 kHz is due to relaxation oscillations of the laser. Magenta: Estimated frequency noise of the interrogation oscillator (laser 1), $S_y^{(2)}(f)$. A contribution to the latter is the noise of laser 2 and an estimate for it is the green trace. It is the spectral density of the frequency fluctuations of a beat between two similar cavity-stabilized lasers, one of which is laser 2, normalized to their optical frequency ($\nu^{(2)}/2 \simeq 282$ THz). Blue: total background noise of the detection process, $S_{\text{det}}(f)$, obtained with the probe laser far detuned from the Doppler-free resonance. A contribution to the latter is the detector noise in absence of laser light, shown in cyan.

*Characterization.*

Figure 2a top shows one characteristic feature of the apparatus: the discriminator. It is recorded by slowly scanning the frequency of the laser $\nu^{(2)}$ across the molecular resonance $\nu^{(1)}$ and recording the signal (voltage) change. The full width of the saturated absorption resonance was 3.6 MHz (FWHM), caused by a relatively high iodine pressure and contamination in the iodine cell. The natural linewidth of the transition is approximately 300 kHz. For DM detection, the laser 1 frequency was tuned to the half-height of the resonance by tuning the local oscillator frequency of the phase lock. For the input signal to the DAQ system, the discriminator on the side of the molecular resonance is $D \simeq 1$ V/MHz.

The second characteristic is the noise level of the apparatus. It was determined by recording, for a comparatively short duration, the signal $V(t)$ when $\nu^{(2)}$ is kept at the operating point. (This recording was not part of the long-duration data recording, so noise presented in the following is to be regarded as typical.) This noise $[\hat{S}_{\text{tot}}(f)]^{1/2}$ is

shown in Fig. 3 in **black** and is given by $\hat{S}_{tot}(f) = (S_y^{(1)}(f) + S_y^{(2)}(f) + S_{det}(f))(\nu^{(2)}D)^2$. Here, $S_y^{(1)}(f)$ is the frequency noise spectrum of the molecular transition frequency, $S_y^{(2)}(f)$ is the frequency noise of the interrogating laser, and $S_{det}(f)$ is the detection noise.

The detection noise is due to effects such as laser amplitude noise, detector noise, electronic noise pick-up, optical feedback. These technical noises dominate over the also present fundamental noise sources, quantum fluctuations (shot noise) and thermal noises. The detection noise $(\nu^{(2)}D)^2 S_{det}(f)$ can be estimated from the spectrum of the signal $V(t)$ when the laser frequency is tuned away from the resonance, since then the effects of $S_y^{(1)}(f)$ and $S_y^{(2)}(f)$ do not appear. The detection noise is shown in blue in Fig. 3.

The background detector noise floor, that includes the noise of the 250 kSa/s data acquisition card (DAQ, digitizer), is a contribution to $(\nu^{(2)}D)^2 S_{det}(f)$ and is shown in cyan. It was measured by blocking the laser beams impinging on D1 and D2. By comparison with the blue trace we find it to be negligible.

An estimate of the total frequency noise of the interrogating laser 1, $S_y^{(2)}(f)$, is shown in magenta. We believe that it is negligible compared to the other noises in the experiment. The laser 1 noise was obtained as follows. Laser 1 is phase-locked to laser 2. The frequency noise of laser 2 is determined by a beat against a similar system (laser 3). The beat linewidth is less than 1 Hz. The green trace shows the linear spectral density of frequency noise of the beat. However, because the apparatus comprising laser 2 and laser 3 is located on the same vibration isolation system, there is certain degree of common-mode noise rejection, and therefore the green curve is only an approximation to the true frequency noise of laser 2. To this noise we add the directly measured noise of the phase-lock between laser 1 and laser 2. The result is the magenta trace.

Note that the monolithic Nd:YAG resonator exhibits an acoustic cutoff, $f_2 \simeq v_s/L_r \simeq 300$ kHz, where $L_r \simeq 2$ cm is the linear dimension of the resonator and $v_s \simeq 6 \times 10^3$ m/s is the speed of sound in the YAG resonator material. However, this cut-off is not relevant here because it is higher than the studied frequency range of experiment A.

The iodine transition frequency $\nu^{(1)}$ exhibits fluctuations induced by varying environmental parameters (temperature, pressure, acoustics). These result in $S_y^{(1)}(f)$ to change over time impacting the the discriminator constant. These variations are difficult to quantify. Because we find that $(\nu^{(2)}D)^2 S_{det}(f)$ and $\hat{S}_{tot}(f)$ are of similar magnitude, we can deduce that the contribution of $S_y^{(1)}(f)$ to the overall noise spectrum is not major. Thus, it appears that the major noise source is residual laser intensity noise.

*Data analysis and presentation*

The time series data were cropped to a duration $T$ so that the number of data points $N$ is equal to a power of 2. Here, $N = 2^{34}$, and $T \simeq 19$ h. The data set $V_i$ was Fourier-transformed, i.e. the fast Fourier transform (FFT) $\tilde{V}_k$ was computed (no normalization was included). A rectangular window function was used. One frequency bin corresponds to $\simeq 15$ µHz. The normalized periodogram $P_k = |\tilde{V}_k|^2/N^2$ was then obtained. The power spectral density is $T P_k$.

The data is presented as a filtered spectrum in Fig. 2 of the main text. This spectrum was generated as follows. The overall spectral range from 10 Hz to 100 kHz was divided into intervals $j$ whose limits were defined, in each decade, by an E48-series distribution. The intervals were (10, 10.5) Hz, (10.5, 11.0) Hz,..., (95.3, 100.0) Hz, etc. For each interval $(f_{j,1}, f_{j,2})$ we consider the center frequency $f_{a,i} = (f_{j,1} + f_{j,2})/2$ as axion frequency and define a simplified DM line-shape model $L(f_{a,j}, f)$, a Lorentzian. Its quality factor $Q_0 \simeq 1 \times 10^6$ is independent of $f_{a,j}$ and thus of the considered interval $j$. Note that the full linewidth $f_{a,j}/Q_0$ ranges from $\simeq 10$ µHz at 10 Hz to $\simeq 100$ mHz at 100 kHz. Thus, at the lower end of the spectrum, 10 Hz, the assumed spectral line is effectively only $\simeq 1$ frequency bins wide.

In each interval $j$ the normalized periodogram $\{P_k\}_j$ is convoluted with the corresponding $L(f_{a,j}, f)$, yielding the optimally filtered periodogram (OFP). Its square root (ROFP) is shown, in simplified form, as orange band in Fig. 2 a of the main document. The trend of the mean of ROFP as a function of Fourier frequency $f$ is due to various noise sources, in particular the increase for $f > 10^4$ Hz is due to laser amplitude noise.

Obvious technical noise has been removed from the ROFP by removing a 5 Hz wide frequency range around every multiple of 50 Hz ranging from 100 Hz to 15 kHz. The convolution applied to the data suppresses signals that are not representative of the expected signals of the galactic halo model. The suppression is weaker for the sun and earth halo models, that have higher signal Q factors.

| $f$ (Hz) | FWHM (bin) | Q-factor ($10^6$) | comment |
|---|---|---|---|
| 642 | 3 | 24 | unknown source |
| 1284 | 3 | 48 | second harmonic |
| $n \times 7681$ | 60 to 300 | 8.5 to 22 | laser PLL; harmonics up to $100\,\mathrm{kHz}$ |
| 46091 | 700 | 0.9 | laser PLL |
| 53670 | 26000 | 0.013 | laser PLL |
| 61454 | 700 | 1.2 | laser PLL |

Table I. Some frequency windows containing technical noise. The quality factors Q are computed from the measured full half-widths at half maximum (FWHM).

*Analysis of technical noise: heuristic approach.*

Table I lists omitted frequency ranges that contain technical noise of substantial strength. In the following further analysis, these ranges as well as those around the $50\,\mathrm{Hz}$-noise peak and its harmonics have been omitted.

A second step in the data analysis consists in checking weaker narrow-linewidth signals. We first mark any frequency bin having a signal strength above the detection limit. The latter is heuristically chosen as 3 standard deviations of the ROFP above its mean. Both standard deviation and mean are computed from the ROFP data in a small spectral window around the frequency of interest. Frequency bins that are above the threshold but within the expected lineshape of a previously marked bin will be rejected to avoid double counting. After the this procedure no candidate remained

*Determination of detection limits: generalities*

The value of the DM field $\phi$ in SI units, at the location of the experiment, can be expressed as [3–5]

$$\phi(t) = \frac{\phi_0}{2\pi f_\phi} \sum_i \alpha_i F'(\omega_i) \cos(\omega_i t + \varphi_i),$$ (1)

where $f_\phi$ is the Compton frequency of the DM particle, $\phi_0 = \sqrt{4\pi \rho_{\mathrm{DM}} G_{\mathrm{N}}/c^2}$ is a normalized field amplitude, $\rho_{\mathrm{DM}}$ is the local DM energy density, and $G_{\mathrm{N}}$ is the gravitational constant. The dimensionless amplitudes $\alpha_i$ (or order unity) and phases $\varphi_i$ are random numbers drawn from specific probability distributions [4]. $\{\omega_i\}$ are a set of regularly spaced angular frequencies starting at $2\pi f_\phi$ and extending over a narrow spectral window. Finally, $F'$ is a dimensionless weighting function that takes into account the velocity distribution of the DM particles and is specific to the assumed DM model. In particular, it is characterized by a fractional full-width at half maximum, $Q^{-1} \ll 1$ (see further below). The DM spectrum is proportional to $F'(\omega)^2$. In the galactic halo DM model, $\rho_{\mathrm{DM}}^{(\mathrm{G})} \simeq 0.3\,\mathrm{GeV/cm^3}$ and thus $\phi_0 \simeq 7 \times 10^{-16}\,\mathrm{Hz}$.

*Determination of detection limits: approach of Derevianko (2018).*

For evaluation of experiment A, we apply Eq. (14) of Ref. [5]. Our 95% bound is taken as $2\hat{\sigma}_{\gamma X}^{(1)}/\Delta R_g^{(\mathrm{A})}$, accounting for the experiment-specific sensitivities.

For the Earth halo model with its infinite coherence time, there is but a single amplitude, $\alpha_1 = \sqrt{2}$, and the bound has a particularly simple expression.

In the galactic and Sun halo models, the above bound holds for observations times significantly longer than the coherence time, say $T \geq 10\,\tau_{\mathrm{coh}}$. For those low DM frequencies $f_\phi$ for which the observation time is significantly shorter than the coherence time, $T < 0.1\,\tau_{\mathrm{coh}}$, we multiply the above expression by a factor 3, as discussed by Centers et al.[6].

In the intermediate range of observation times, $0.1\,\tau_{\mathrm{coh}} \leq T \leq 10\,\tau_{\mathrm{coh}}$, the theory of the bound has not yet been worked out. We therefore conservatively apply the factor of 3 also in this range.

The following correction to the formulae in Ref. [5] is implemented [7]: in Sec.II A, the factors "2" appearing in the definitions of $d_e$, $d_{m_e}$ are replaced by "4". For the galactic halo model, we use the velocity values given by Foster *et al.* [4], $\sqrt{2}v_{\mathrm{vir}} = v_0 \simeq 220\,\mathrm{km/s}$, $v_g = v_{\mathrm{obs}} \simeq 232\,\mathrm{km/s}$.

## C.   Experiment B

*Apparatus.* The apparatus of experiment B implements Doppler-broadened absorption spectroscopy in the R(122) 2-10 I$_2$ transition at 725 nm, and is shown in Fig. 1b. Iodine vapor is excited with ≈2 mW of light from a Ti:Sapphire laser (M squared SolsTiS) in a 10-cm long cell. The light beam is double-passed through the cell to increase the absorption signal. The cell body is maintained at ≈405 °C, sufficiently high to obtain adequate population in the electronic ground state's vibrational level with $v$=10. The pressure in the I$_2$ cell is set to ≈30 mbar via heating of the cold finger of the cell, that is maintained to 87 °C to within ±1 °C. Balanced detection of the light transmitted through the cell is done with use of a secondary reference beam, to minimize the effects of laser amplitude noise. Small drifts in this balancing are corrected by monitoring the output of the balanced detector (Thorlabs PBD415A) and applying feedback to a stepper-motor mounted HWP (HWP 3 in Fig. 1b) to adjust the power of the beam headed to the I$_2$ cell. This results in suppression of laser amplitude noise by more than ×100 times. The output of the balanced photodetector is amplified ×100 times with a preamplifier (Femto HVA-200M-40-B) and is recorded with a 12-bit DAQ system (Picoscope 5244D) at a rate of 250 MSa/s. An electro-optic modulator (EOM) is used in auxiliary experiments to impose frequency modulation on the laser light, in order to measure the frequency response of the apparatus. This response may be characterized by an overall calibration function $h(f)$, which is primarily determined by the decaying response of molecules at frequencies larger than the transition's linewidth. This frequency modulation is checked with a Fabry-Perot (FP) cavity whose resonance has ≈ 150 MHz FWHM. The peak I$_2$ absorption corresponds to ≈ 1 absorption length for the Doppler- and pressure-broadened resonance The transition width of ≈ 970 MHz has contributions due to Doppler (≈ 485 MHz) and collisional broadening (≈ 240 MHz). The latter is estimated from comparison of the 970 MHz width with the value ≈ 730 MHz observed at much lower I$_2$ pressure (≈ 3 mbar). The molecular response is essentially constant over the 100 MHz range probed for FC oscillations (the function $h(f)$ was measured to be ≃ 1 for all frequencies probed).

*Experimental protocol.*
An experimental run proceeds as follows: First, the laser frequency is swept over the I$_2$ resonance and the reference beam power is set to obtain a zero-crossing of the photodetector output on the side of the resonance, at the half-height of absorption feature (see Fig. 2b). Then the laser frequency is tuned to the nominal zero crossing and stabilized to the reading of a He-Ne referenced wavemeter (drift ≈ 2 MHz/h), and subsequent slow drifts of the detector output are actively compensated, as mentioned above. A time series of the amplified detector signal is recorded in a 0.1 s window once every ≈1 s (i.e. 10% measurement duty cycle), and corresponding FFT of these data are computed and the resulting periodograms, i.e. the squared magnitudes of the computed periodograms, are continuously averaged. A flattop window is applied to the time-series data, to avoid parasitic effects in the periodograms due to the discrete nature of the computation. This windowing leads to an effective broadening of the single-bin width in the frequency domain to a resulting ≈ 37.7 Hz (It additionally reduces the integration time from 0.1 s to an effective 27 ms.). Thus, the resulting periodogram in the 100 kHz-100 MHz range consists of $N \approx 2.65 \cdot 10^6$ bins. After ≈ 200 s of data taken on the slope of the resonance, the laser frequency is detuned from resonance by 1.1 GHz, where the discriminator slope is ≈ 0, and the power of the beam headed to the I$_2$ cell is re-adjusted with feedback to maintain balanced detection, as mentioned above. Equal amount of data as before are taken in this FC-oscillation insensitive configuration. The corresponding periodogram is subtracted from that from data with sensitivity to FC oscillations, to minimize the impact of parasitic effects such as laser amplitude noise and instrumentation pickup. After cycling many times between the two configurations, a difference periodogram is obtained that is nearly free of parasitics and will contain power in excess of noise in the presence of FC oscillations. We henceforth refer to this as excess power spectrum (PS). We show this excess PS, produced using the data of our main 60-h-long DM run in Fig. 4.

Note that this total measurement time exceeds the coherence times of the galactic halo and solar halo models for the considered range of $f_\phi$ values, so that the Rayleigh probability distribution of the amplitudes $\alpha_i$ is fully sampled.

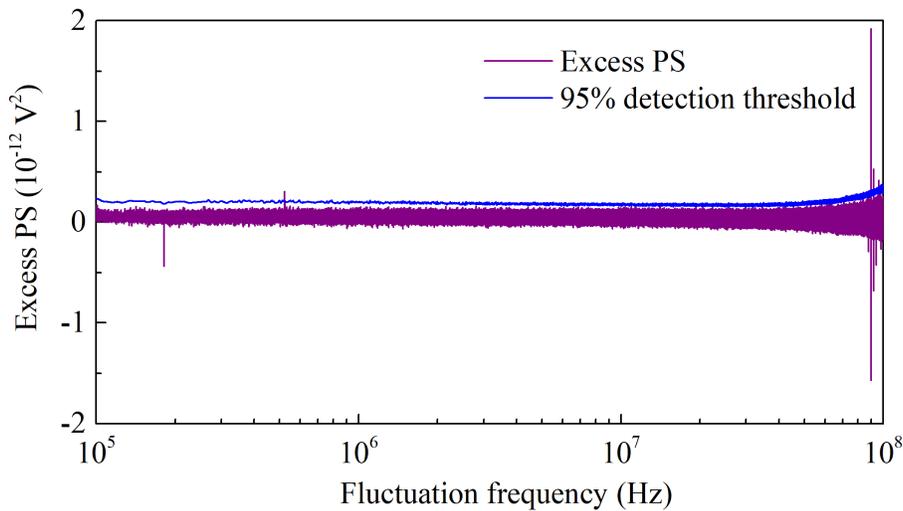

Figure 4. Excess PS and detection threshold at the 95% CL.

*Check for DM candidate signals.*

The excess PS of Fig. 4 is analyzed for possible FC oscillations. A DM signature would be excess power in the spectrum, above a detection threshold computed below and shown in Fig. 4. In addition to investigating such candidate signals, 'negative' peaks (i.e. with power significantly smaller than the mean background noise) are also investigated. It was established that all such peaks have technical origins and although their power in most cases is expected to cancel out in the excess PS, residual power for some of those remains. Checking the residual background power after they are accounted for, informs about potential FC oscillations in the respective frequency positions, as is the case of candidates with high excess power.

Auxiliary experiments were carried out to identify the origin of the candidate DM and spurious peaks and measure their respective powers. These experiments were done with use of a calibrated spectrum analyzer (Keysight N9320B) to acquire spectra over 5 kHz regions around the candidate peaks. This device is more time efficient in acquiring spectra than our primary acquisition system when recording spectra over narrow frequency windows. In one set of measurements, data were acquired to check for the high-frequency-range spurious signals (above 60 MHz), alternating between acquisition with laser frequency tuned on and off the $I_2$ resonance. The new excess PS was found to be consistent with background noise. All spurious signals above 60 MHz were found to be due to rf apparatus pickup, since the corresponding peaks were present in the absence of light arriving at the detector. In another experiment, an EOM (EOM 2 in Fig. 1b) was used to provide improved suppression of amplitude noise in the balanced photodetection employed in the setup. This resulted in the elimination of a peak at $\approx 181$ kHz, which was identified to be due to laser amplitude noise, as it appeared in direct measurement of the laser output light. Finally, in another experiment, the power of a peak at $\approx 523$ kHz, only present with the laser tuned on the slope of the $I_2$ resonance, was identified to be laser frequency noise with the FP (see Fig. 1b). Its power was measured on the slope of the FP resonance and subtracted from the excess PS. As a result of all checks following the main 60-hr long DM run, it was established that all peaks under investigation had a technical source, and after these were accounted for in the excess PS, the residual power in the respective frequencies was consistent with background noise and below the detection threshold.

*Computation of DM constraints.*

Detection of FC oscillations in the excess PS of Fig. 4 is associated with power higher than noise. This noise is used to define a detection threshold $P_{th}$ at the 95% confidence level, so that if a spectral feature has excess power $P_{ex} > P_{th}$, there is probability $p_0 = 5\%$ that it is due to statistical fluctuations. Given a noise distribution function and associated cummulative distribution function (CDF), one can express $p_0$ as:

$$p_0 = 1 - \mathrm{CDF}(P_{th}), \tag{2}$$

where $\mathrm{CDF}(P_{th})$ represents the probability $p(P_{ex} < P_{th})$ that the excess power is smaller than the threshold. The Eq. (2) may be be solved for the detection threshold $P_{th}$. However this determination would only be valid for an experiment investigating a single frequency bin. In experiment B, where FC oscillations are looked for in $N \approx 2.65 \times 10^6$ bins ($\approx 100$ MHz search window consisting of 37.7 Hz bins), the threshold has to be raised to account for the fact that a fraction of bins ($\approx 5\%$) are expected to have power in excess of $P_{th}$ [8]. This requires raising the CDF in Eq. (2) in

the $N^{th}$ power, so that:

$$p_0 = 1 - \big[\text{CDF}(P_{\text{th}})\big]^N. \tag{3}$$

The noise distribution of the excess PS of Fig. 4 was checked in many frequency bins and it was found to be well described by a Gaussian. Given this, one can solve for the threshold power $P_{\text{th}}$ using the expression for the Gaussian CDF, and obtain:

$$P_{\text{th}} = \sqrt{2}\text{erf}^{-1}\Big\{2(1-p_0)^{1/N} - 1\Big\}\sigma + \mu \approx 5.5\sigma + \mu, \tag{4}$$

where $\mu$ and $\sigma$ are the mean and standard deviation of the Gaussian noise. These parameters were determined throughout the spectrum of Fig. 4 by fitting the noise in consecutive 5-kHz-wide windows.

From the determined threshold $P_{\text{th}}$, constraints are extracted on the frequency fluctuation $\delta\nu$ . This fluctuation is given by $\delta\nu = \delta V/D$, where $\delta V$ is the voltage fluctuation corresponding to the fluctuation $\delta P$ in the excess PS spectrum of Fig. 4 (at the 95% confidence level: $\delta P = P_{\text{th}} - \mu = 5.5\sigma$). The discriminator slope $D$ is discussed in the main text. The quantities $\delta V$ and $\delta P$ are related via $\delta P = 2V_{\text{av}}\delta V = 2\sqrt{P_{\text{av}}}\delta V$, where $P_{\text{av}} = V_{\text{av}}^2$ is the averaged PS recorded with sensitivity to FC oscillations (i.e. with laser tuned to the slope of the $I_2$ resonance). One obtains for $\delta\nu$:

$$\delta\nu = \frac{\delta V}{D} = \frac{\delta P}{2\sqrt{P_{\text{av}}}D} = \frac{5.5\sigma}{2\sqrt{P_{\text{av}}}D}. \tag{5}$$

### Effect of decoherence of DM field.

Before placing constraints on FC, one has to consider the effects of partial decoherence of the DM field, that if present, will result in reduced sensitivity to DM detection. In experiment B, decoherence needs to be accounted for within the galactic DM halo scenario: $\tau_{\text{coh}} \simeq (6, 0.006)\,\text{s}$ at $f_\phi = (10^5, 10^8)\,\text{Hz}$. Within the Solar and Earth halo scenarios, the Q-factor of the field is high enough so that there is negligible decoherence over the 27 ms acquisition time. In practice, a sensitivity penalty must be applied to the obtained $\delta\nu/\nu$ spectrum, to account for the decoherence during the effective 27-ms-long time interval of our data acquisition. This penalty becomes significant at a frequency $f_\phi \geq 1/27\,\text{ms} \approx 40\,\text{MHz}$. To compute this sensitivity loss, we considered the lineshape, $f_{\text{DM}}(f)$, that arises in the lab as the laboratory moves through the virialized DM field with a velocity dispersion of $v_0 \approx 10^{-3}c_0$ (denoted above by $\sqrt{2}v_{\text{vir}}$),

$$f_{\text{DM}}(f) = \frac{2c^2}{\sqrt{\pi}v_0 v_{\text{lab}} f_\phi}\exp\left(-\frac{2c^2}{v_0^2}\frac{f - f_\phi}{f_\phi} - \frac{v_{\text{lab}}^2}{v_0^2}\right)\sinh\beta, \tag{6}$$

where $v_{\text{lab}}$ (denoted $v_g$ above) is the velocity of the laboratory in the galactic frame (232 km/s), and we have denoted

$$\beta = \frac{2c v_{\text{lab}}}{v_0^2}\sqrt{\frac{2(f - f_\phi)}{f_\phi}} \tag{7}$$

for brevity. This lineshape is proportional to the power spectral density of the DM particle and has been derived previously in Refs. [4, 5, 9, 10]. (The relationship to $F'$ is $f_{\text{DM}}(f) = F'(\omega)^2 T$.)

We had a constant frequency bin width equal to 37.7 Hz in experiment B due to the flat-top windowing in the time domain. At frequencies higher than $\approx 40\,\text{MHz}$ the DM power spectral density will be broader than that of the bin-width, effectively leading to the power spreading over more than one point in our spectrum. We have calculated the loss due to this for different DM particle Compton frequencies, $f_\phi$, by integrating the lineshape in 37.7 Hz bins and steps of 10 Hz. The maximum of these bin-integrals is then compared to the total area of the lineshape. This procedure yields the fraction of the DM power that will be observed in our spectrum. For a sense of scale, at 10 MHz the sensitivity loss because of this is $\approx 1\%$ and at 100 MHz it is $\approx 75\%$.

## II.  THEORY

*Details of the Earth halo model.*

This model may alternatively be called the "gravitational hydrogen atom model". The DM field is monochromatic with infinite coherence time, following from the assumption that the earth halo is infinitely stable, i.e. the virial velocity is zero. The value of $\rho_{\mathrm{DM}}^{\oplus}$ is a function of DM particle mass (see Fig. 2a in supplementary information of Ref. [11]), and is enhanced compared to $\rho_{\mathrm{DM}}^{\mathrm{G}}$ by a factor increasing from $10^4$ at $f_\phi = 100\,\mathrm{Hz}$ to $10^{19}$ at $f_\phi = 3.4\,\mathrm{MHz}$. However, beyond $f_\phi \simeq 15\,\mathrm{MHz}$ the ratio $\rho_{\mathrm{DM}}^{\oplus}/\rho_{\mathrm{DM}}$ drops below 1.

*Equivalence Principle (EP) tests.*

A scalar field $\phi$ with mass $m_\phi$ would induce an Yukawa interaction between two bodies A and B in the range $\Lambda = h/m_\phi c$ which is non-universal and thus violates the equivalence principle (EP). The total potential (gravitational and Yukawa) between these two bodies can be written as

$$V = -G_{\mathrm{N}} \frac{m^{\mathrm{A}} m^{\mathrm{B}}}{r_{\mathrm{AB}}} \left( 1 + \alpha^{\mathrm{A}} \alpha^{\mathrm{B}} e^{-r_{\mathrm{AB}}/\Lambda} \right) , \qquad (8)$$

where, $m^{\mathrm{A}}$ ($m^{\mathrm{B}}$) is the mass of body A (B), $G_{\mathrm{N}}$ is the Newtonian gravitational constant and $r_{\mathrm{AB}}$ is the distance between A and B. $\alpha^{\mathrm{A,B}}$ are the strengths of the Yukawa interaction. They measure the susceptibility of the mass to $\phi$, and thus can be written as,

$$\alpha = \frac{1}{\sqrt{4\pi G_{\mathrm{N}}}} \frac{\partial \ln m(\phi)}{\partial \phi} . \qquad (9)$$

In the presence of a central body (the "source") S with mass $m^{\mathrm{S}}$ at a distance $r$, the acceleration of a test body A can be written as

$$\vec{a}_{\mathrm{A}} = -\hat{r} \frac{G_{\mathrm{N}} m^{\mathrm{S}}}{r^2} \left[ 1 + \alpha^{\mathrm{A}} \alpha^{\mathrm{S}} \left( 1 + \frac{r}{\Lambda} \right) e^{-r/\Lambda} \right] . \qquad (10)$$

The Eötvös parameter, $\eta_{\mathrm{EP}}$, which measures the differential acceleration between two test bodies A and B in the presence of a source S, follows as as [12, 13]

$$\eta_{\mathrm{EP}}^{\mathrm{Exp}} \equiv 2 \frac{|\vec{a}_{\mathrm{A}} - \vec{a}_{\mathrm{B}}|}{|\vec{a}_{\mathrm{A}} + \vec{a}_{\mathrm{B}}|} \simeq (\alpha^{\mathrm{A}} - \alpha^{\mathrm{B}}) \alpha^{\mathrm{S}} \left( 1 + \frac{r}{\Lambda} \right) e^{-r/\Lambda} . \qquad (11)$$

EP test experiments constrain $\eta_{\mathrm{EP}}^{\mathrm{Exp}}$ as a function of $\Lambda$. This leads to bounds on $\alpha^{\mathrm{A,B,S}}$ as a function of $m_\phi$. Because $\alpha^{\mathrm{A,B,S}}$ depend on the fundamental constants (FCs) as shown in the following, a bound on $\eta_{\mathrm{EP}}$ can be converted into a bound on the coupling coefficients $d_i$ .

*FC dependence of atom mass.*

To discuss how the mass of a body depends on FCs, let us start by noting that the mass of a generic atom a with atomic number $A^{\mathrm{a}}$ and proton number $Z^{\mathrm{a}}$ can be expressed as,

$$m^{\mathrm{a}} = m_N^{\mathrm{a}}(A^{\mathrm{a}}, Z^{\mathrm{a}}) + Z^{\mathrm{a}} m_e . \qquad (12)$$

where $m_N^{\mathrm{a}}$ is the mass of the nucleus of atom a and $m_e$ is the electron mass. The mass of the nucleus can be further decomposed as

$$m_N^{\mathrm{a}}(A^{\mathrm{a}}, Z^{\mathrm{a}}) = Z^{\mathrm{a}} m_p + (A^{\mathrm{a}} - Z^{\mathrm{a}}) m_n + E_3 + E_1 , \qquad (13)$$

where, $m_p$ ($m_n$) is the proton (neutron) mass and $E_3$ ($E_1$) is the binding energy of the strong (electromagnetic) interaction. Note that $E_1$ is dominated by the electromagnetic effect within the nucleus [14] and thus we will ignore the electron effect on this. As $m_p, m_n, E_3$ and $E_1$ depend on the FCs, variation of the FCs would lead to a variation of nucleus and atom mass, and can be written as

$$\frac{\partial \ln m^{\mathrm{a}}(\phi)}{\partial \phi} = \frac{\partial \ln m^{\mathrm{a}}(\phi)}{\partial \ln g_i} \frac{\partial \ln g_i}{\partial \phi} = Q_i^{\mathrm{a}} \frac{d_i}{M_{\mathrm{Pl}}} , \qquad (14)$$

where $g_i$ is a generic FC, and we have also implied summation over repeated indices here and below. We have introduced the notation $Q_i^{\mathrm{a}} \equiv \partial \ln m^{\mathrm{a}}(\phi)/\partial \ln g_i$, the $i$th "dilatonic charge" of a body. The susceptibility of a FC to $\phi$ can be rewritten as $\partial \ln g_i/\partial \phi = d_i/M_{\mathrm{Pl}}$ (See Eqs. (2-5) in the main text).

For a given body, the dilatonic charges corresponding to different FCs can be given in vector form explicitly as [14] (with $(\vec{Q}^{\mathrm{a}})_i = Q_i^{\mathrm{a}}$),

$$\vec{Q}^{\mathrm{a}} \approx F^{\mathrm{a}}\Big(3 \times 10^{-4} - 4\,r_I + 9\,r_Z,\, 3 \times 10^{-4} - 3\,r_I,\, 1,$$
$$0.1 - \frac{0.04}{(A^{\mathrm{a}})^{1/3}} - 2 \times 10^6 r_I^2 - r_Z,\, 0.002\,r_I\Big),\tag{15}$$

where here and below $\vec{X} \equiv (X_\alpha, X_{m_e}, X_{g_s}, X_{\hat{m}}, X_{\delta m})$, with $\hat{m}, \delta m/2 \equiv (m_u \pm m_d)/2$,

$10^4\,r_I \equiv 1 - 2Z^{\mathrm{a}}/A^{\mathrm{a}}$, $10^4\,r_Z \equiv Z^{\mathrm{a}}(Z^{\mathrm{a}}-1)/(A^{\mathrm{a}})^{4/3}$, and $F^{\mathrm{a}} = 931\,A^{\mathrm{a}}/(m^{\mathrm{a}}/\mathrm{MeV})$.

Finally, the Yukawa strength of a body composed of atoms of species a is $\alpha^{\mathrm{a}} = \sqrt{2}\,\vec{Q}^{\mathrm{a}} \cdot \vec{d}$, where we have used $M_{\mathrm{Pl}} = \sqrt{\hbar c/(8\pi G_{\mathrm{N}})}$.

As we discussed in the main text in the context of EP tests, and in the following here, we do not consider the strange quark mass contribution.

For our experiment we can also construct an analogous $\vec{Q}$, setting $Q_i^{\mathrm{X}} = \Delta R_i^{(\mathrm{X})}$ with $\mathrm{X = A, B}$ as given in the main text, with a relative sensitivity to the QCD parameters of 0.06 (-0.07) in experiment A (B). So,

$$\vec{Q}^{\mathrm{A}} \simeq \big(1, 0.06, -0.05, -0.005, -2 \times 10^{-5}\big),$$
$$\vec{Q}^{\mathrm{B}} \simeq \big(2, 0.93, 0.06, 0.006, 2 \times 10^{-5}\big).\tag{16}$$

*Sensitivities of EP tests.*

Let us now discuss the sensitivities of the EP test experiments. In the mass region of our interest $10^{-14}\,\mathrm{eV} \lesssim m_\phi \lesssim 10^{-6}\,\mathrm{eV}$ these tests have given stronger bounds than those arising from the direct scalar DM searches. However, as we have discussed in the main text and also here, the EP tests compare the dilatonic charges of two test bodies. For instance, consider the MICROSCOPE experiment [15, 16], that provides one of the strongest EP bounds for masses below $10^{-12}\,\mathrm{eV}$. It is sensitive to the difference of the dilatonic charges of a platinum/rhodium alloy (90%/10%) and a titanium/aluminum/vanadium alloy (90%/6%/4%), with $(\overrightarrow{\Delta Q})^{\mathrm{Mic}} \approx 10^{-3}(-1.94,\, 0.03,\, 0.8,\, -2.61,\, -0.19)$ [14]. Alloys are treated as mixtures of pure systems. By inspecting the enclosed angles, we find that the direction of this vector in coupling sapce is very different than that of our experiments $\vec{Q}^{\mathrm{A,B}}$.

In our present discussion, we consider the six most sensitive EP test experiments. Apart from MICROSCOPE, these are [12, 17–20]:

$$(\overrightarrow{\Delta Q})^{\mathrm{Be-Al}} \simeq 10^{-3}(-1.021,\, -0.021,\, -2.035,\, -5.498,\, 0.126),$$
$$(\overrightarrow{\Delta Q})^{\mathrm{Be-Ti}} \simeq 10^{-3}(-1.560,\, -0.008,\, -2.435,\, -7.412,\, 0.047),$$
$$(\overrightarrow{\Delta Q})^{\mathrm{Be-Cu}} \simeq 10^{-3}(-1.982,\, -0.007,\, -2.462,\, -8.204,\, 0.039),$$
$$(\overrightarrow{\Delta Q})^{\mathrm{Cu-Pb(a)}} \simeq 10^{-3}(-1.564,\, 0.033,\, 0.945,\, -1.764,\, -0.203),$$
$$(\overrightarrow{\Delta Q})^{\mathrm{Cu-Pb(b)}} \simeq 10^{-3}(-1.412,\, 0.030,\, 0.951,\, -1.427,\, -0.185),$$

where $\mathrm{Pb^{(a)}}$ denotes the alloy of composition lead/antimony/tin (92%/7.75%/2.5%) [18] and $\mathrm{Pb^{(b)}}$ denotes the alloy of composition lead/aluminium (91.5%/8.5%) [20]. Note that we have denoted the $\mathrm{Pb^{(a)}}$ combination as Pb and $\mathrm{Pb^{(b)}}$ combination as Pb-alloy in the main text.

*Finding directions in coupling space to which EP tests are weakly sensitive to.*

In the five dimensional vector space of couplings, we can construct a unit vector orthogonal to four most stringent EP tests for each mass. For example, in the mass range of $2 \times 10^{-12} \lesssim m_\phi/\mathrm{eV} \lesssim 5 \times 10^{-9}$, the leading EP tests are Be-Ti, Be-Al, Be-Cu and Cu-Pb and in this mass range the orthogonal vector can be given as,

$$\hat{Q}_{\mathrm{Full}}^{\perp} \simeq \big(0.003,\, -0.987,\, 0.002,\, -0.001,\, -0.162\big).$$

Its existence implies that models of light scalar DM with this combination of relative coupling amplitudes are not constrained by these four leading EP test experiments. Now, $\hat{Q}_{\mathrm{Full}}^{\perp}$ has a large overlap with the $d_{m_e}$ direction (the 2nd entry of $\hat{Q}_{\mathrm{Full}}^{\perp}$). Thus, experiments that are sensitive to time-variation of the electron mass are sensitive to a sector of DM - SM coupling parameter space that the first four-best EP bounds are insensitive to, barring a coincidence. For our experiments, $\hat{Q}_{\mathrm{Full}}^{\perp} \cdot \vec{Q}^{\mathrm{A}} \simeq -0.03$, and $\hat{Q}_{\mathrm{Full}}^{\perp} \cdot \vec{Q}^{\mathrm{B}} \simeq -0.87$; these values are approximately the sensitivity coefficients for $d_{m_e}$. The sensitivity of our current experiments into the $\hat{Q}_{\mathrm{Full}}^{\perp}$ direction is indicated by the dotted

lines in Fig. 5 in the main text (further projected onto the $d_{m_e}$ direction for ease of display). The brown dotted line depicts the bound of the remaining 5th-best EP experiment. In the $\hat{Q}^{\perp}_{\text{Full}}$ direction EP tests constraints are stronger than our experiments by 2-3 orders of magnitude, compared to only $d_{m_e} \neq 0$ models, where the EP tests are stronger by 8-10 orders of magnitude.


\fi


\end{document}